\pdfoutput=1
\RequirePackage{ifpdf}
\ifpdf 
\documentclass[pdftex]{sigma}
\else
\documentclass{sigma}
\fi

\numberwithin{equation}{section}

\usepackage[all,cmtip]{xy}

\usepackage{tikz}
 \usetikzlibrary{automata,positioning,arrows}
\usepackage{filecontents}

\tikzset{
 net node/.style = {draw, circle, minimum size=8mm},
 net edge/.style = {->, >=triangle 45}, 
 net cut/.style = {shorten >=-10mm, shorten <=-10mm, rounded corners=10mm, color=red},
 net cross/.style = {sloped, allow upside down, pos=.3},
}

\begin{document}

\newcommand{\arXivNumber}{2010.15276}

\renewcommand{\PaperNumber}{005}

\FirstPageHeading

\ShortArticleName{Ladder Operators and Hidden Algebras for Shape Invariant.~II}

\ArticleName{Ladder Operators and Hidden Algebras\\ for Shape Invariant Nonseparable\\ and Nondiagonalizable Models
with Quadratic\\ Complex Interaction. II.~Three-Dimensional Model}

\Author{Ian MARQUETTE~$^{\rm a}$ and Christiane QUESNE~$^{\rm b}$}

\AuthorNameForHeading{I.~Marquette and C.~Quesne}

\Address{$^{\rm a)}$~School of Mathematics and Physics, The University of Queensland,\\
\hphantom{$^{\rm a)}$}~Brisbane, QLD 4072, Australia}
\EmailD{\href{mailto:i.marquette@uq.edu.au}{i.marquette@uq.edu.au}}

\Address{$^{\rm b)}$~Physique Nucl\'eaire Th\'eorique et Physique Math\'ematique,
Universit\'e Libre de Bruxelles,\\
\hphantom{$^{\rm b)}$}~Campus de la Plaine CP229, Boulevard~du Triomphe, B-1050 Brussels, Belgium}
\EmailD{\href{mailto:christiane.quesne@ulb.be}{christiane.quesne@ulb.be}}

\ArticleDates{Received September 01, 2021, in final form January 03, 2022; Published online January 14, 2022}

\Abstract{A shape invariant nonseparable and nondiagonalizable three-dimensional model with quadratic complex interaction was introduced by Bardavelidze, Cannata, Ioffe, and Nishnianidze. However, the complete hidden symmetry algebra and the description of the associated states that form Jordan blocks remained to be studied. We present a set of six operators $\{A^{\pm},B^{\pm},C^{\pm}\}$ that can be combined to build a ${\mathfrak{gl}}(3)$ hidden algebra. The latter can be embedded in an ${\mathfrak{sp}}(6)$ algebra, as well as in an ${\mathfrak{osp}}(1/6)$ superalgebra. The states associated with the eigenstates and making Jordan blocks are induced in different ways by combinations of operators acting on the ground state. We present the action of these operators and study the construction of an extended biorthogonal basis. These rely on establishing various nontrivial polynomial and commutator identities. We also make a~connection between the hidden symmetry and the underlying superintegrability property of the model. Interestingly, the integrals generate a cubic algebra. This work demonstrates how various concepts that have been applied widely to Hermitian Hamiltonians, such as hidden symmetries, superintegrability, and ladder operators, extend to the pseudo-Hermitian case with many differences.}


\Classification{81Q05; 81Q60; 81R12; 81R15}

\section{Introduction}

Over the years, a large body of literature has been devoted to non-Hermitian Hamiltonians with a~real spectrum, in particular PT-symmetric systems \cite{bender05,bender07,bender98} or, more generally, pseudo-Hermitian ones, for which $\eta H \eta^{-1} = H^{\dagger}$ with $\eta$ a Hermitian invertible operator \cite{mosta02a,mosta10}. The concept of pseudo-Hermiticity was introduced a long time ago by Pauli as generalized Hermiticity~\cite{pauli43} and later on by Scholtz, Geyer, and Hahne as quasi-Hermiticity \cite{scholt92}.

However, mainly the one-dimensional case has been explored and only a few two- and three-dimensional systems have been obtained (see, e.g., \cite{barda,bender01,cannata10,ioffe,nana}) and their study may still be incomplete.

In the previous paper of this series \cite{marquette20}, we re-examined a non-Hermitian two-dimensional system \cite{cannata10} which is exactly solvable due to its shape invariance \cite{bagchi,cannata02,cooper,genden,junker}, although it is neither separable nor diagonalizable. It was demonstrated that this model has a~${\mathfrak{gl}}(2)$ hidden symmetry algebra and that an underlying ${\mathfrak{sp}}(4)$ algebra \cite{mosh} can be constructed, as well as an~${\mathfrak{osp}}(1/4)$ superalgebra~\cite{frappat}. In contrast with the two-dimensional Hermitian harmonic oscillator, however, the algebraic structure is related to integrability, but not superintegrability.

The main purpose of the present paper is to consider another non-Hermitian system with a~quadratic complex interaction, which is exactly solvable, although not separable nor diagonalizable~\cite{barda}. This system being in three dimensions has a more complicated structure of the Jordan blocks that are needed to form a complete basis. This is the reason it deserves a separate study.\looseness=-1

Another purpose of this paper is to highlight properties of non-Hermitian systems in higher dimensions, such as exact solvability without separation of variables and hidden symmetry that does not lead to superintegrability in the usual way. We would also like to emphasize that the ladder operators constructed from partial differential operators may have distinct properties, such as the possibility of infinitely zero modes.

The paper is organized as follows. In Section~\ref{section2}, we review the Hamiltonian of the three-dimensional nonseparable oscillator of \cite{barda}, as well as its spectrum and wavefunctions $\Psi_{k,n,0}(\bf{x})$, and we present the set of known ladder operators $A^{\pm}$, $Q^{\pm}$, which do not close in a finite-dimensional Lie algebra. In Section~\ref{section3}, we construct two additional sets of ladder operators, $B^{\pm}$ and $C^{\pm}$, motivated by their action on the wavefunctions with $k=0$. In Section~\ref{section4}, we use them to build a set of nine bilinear operators satisfying a nine-dimensional Lie algebra, which can be transformed into ${\mathfrak{gl}}(3)$ and proved to provide the Hamiltonian hidden symmetry algebra. A set of bosonic operators in a nonstandard realization is constructed and used to embed ${\mathfrak{gl}}(3)$ in an ${\mathfrak{sp}}(6)$ algebra and an ${\mathfrak{osp}}(1/6)$ superalgebra. The integrals of motion of~\cite{barda} are also interpreted in this context to relate the superintegrability of the model to the hidden symmetry algebra. In Section~\ref{section5}, we demonstrate how we can induce the $\Psi_{k,n,m}({\bf x})$ associated functions in different algebraic ways. We also present the action of the ladder operators and of the ${\mathfrak{gl}}(3)$ linear Casimir operator on all the states belonging to Jordan blocks. The construction of an extended biorthogonal basis \cite{mosta02b, mosta02c} is then discussed in Section~\ref{section6}. Finally, Section~\ref{section7} contains the conclusion.

\section{Shape invariant model with quadratic complex interaction}\label{section2}

Let us consider the three-dimensional model with complex oscillator Hamiltonian \cite{barda}
\begin{gather*}
 H=-\partial_1^2 - \partial_2^2 -\partial_3^2 + \lambda^2\big(x_1^2+x_2^2+x_3^2\big) + g^2
 \big(x_1^2-2{\rm i}x_1x_2-x_2^2\big) - 4 \lambda g (x_1-{\rm i}x_2) x_3 -3 \lambda,
\end{gather*}
where $g$ and $\lambda$ are two real parameters such that $\lambda > |g|$. This Hamiltonian can be rewritten as
\begin{gather}
 H = -4 \partial_z \partial_{\bar{z}} - \partial_3^2 + \lambda^2 \big(z \bar{z} +x_3^2\big) + g^2 \bar{z}^2 - 4
 \lambda g \bar{z} x_3 -3 \lambda, \label{eq:H}
\end{gather}
where $z=x_1 +{\rm i} x_2$ and $\bar{z}=x_1-{\rm i} x_2$. It satisfies pseudo-Hermiticity with $\eta$ chosen as $P_2$, which is the operator changing $x_2$ into $-x_2$.

With the operators
\begin{gather}
 A^{\pm}=2 \partial_z \mp \lambda \bar{z}. \label{eq:A}
\end{gather}
the Hamiltonian (\ref{eq:H}) satisfies the intertwining properties
\begin{gather*}
 HA^{+}=A^{+}(H+2 \lambda ), \qquad A^{-}H=(H+2\lambda)A^{-},
\end{gather*}
showing its self-isospectral shape invariance. These equations can also be expressed in the form of ladder-type relations
\begin{gather}
 [H,A^{\pm}]=\pm 2 \lambda A^{\pm}. \label{eq:ha}
\end{gather}
However, unlike ladder operators for the Hermitian harmonic oscillator, the operators (\ref{eq:A}) satisfy the relation
\begin{gather}
 [A^{-},A^{+}]=0. \label{eq:aa}
\end{gather}
This algebra can be seen as a Euclidean e(2), which differs from the Heisenberg algebra encountered in the Hermitian case.

Another set of ladder operators can be introduced \cite{barda},
\begin{gather}
 Q^{\pm} = 4 \partial_z \partial_{\bar{z}} -\partial_3^2 \mp 2 \lambda ( z \partial_z + \bar{z}
 \partial_{\bar{z}} -x_3 \partial_3 )
 \pm 4 g x_3 \partial_{z} \mp 2 g \bar{z} \partial_3\nonumber\\
 \hphantom{Q^{\pm} =}{} + \lambda^2 \big(z \bar{z} -x_3^2\big) -g^2 \bar{z}^2 \mp
 \lambda. \label{eq:q}
\end{gather}
They satisfy the following commutation relations with $H$,
\begin{gather}
 [H,Q^{\pm}]=\pm 4 \lambda Q^{\pm}, \label{eq:hq}
\end{gather}
which can also be interpreted as intertwining relations, characteristic of shape invariance. Additional commutation relations read
\begin{gather}
 [A^{\pm},Q^{\mp}]=\pm 4 \lambda A^{\pm}, \label{eq:aqd} \\
 [A^{\pm},Q^{\pm}]=0, \label{eq:eqm} \\
 [ Q^{-},Q^{+}]=-2 \tilde{R}_1 =8 \big(\lambda H -2 g R_1 +3 \lambda^2\big), \label{eq:qq}
\end{gather}
where
\begin{gather*}
 R_1=2 \partial_z \partial_3 + \lambda \bar{z} (g \bar{z} -\lambda x_3 ).
\end{gather*}

The Schr\"odinger equation
\begin{gather*}
 H \Psi({\bf x}) = E\Psi({\bf x})
\end{gather*}
can be solved in terms of the operators $A^{\pm}$ and $Q^{\pm}$~\cite{barda}. The (unnormalized) ground-state wavefunction $\Psi_0$, such that
\begin{gather*}
 A^- \Psi_0 = Q^- \Psi_0 = 0,
\end{gather*}
is given by
\begin{gather}
 \Psi_0(z,\bar{z},x_3)={\rm e}^{-\frac{\lambda}{2} (z \bar{z} + x_3^2)+ g \bar{z} x_3} \label{eq:gs}
\end{gather}
with corresponding energy
\begin{gather*}
 E_0=0.
\end{gather*}
The operators $A^+$ and $Q^+$ allow to obtain from it the excited-state wavefunctions
\begin{gather}
 \Psi_{k,n,0}(z,\bar{z},x_3)=c_{k,n} (Q^{+})^k (A^{+})^n \Psi_0(z,\bar{z},x_3), \label{eq:wf}
\end{gather}
corresponding to
\begin{gather*}
 E_{k,n}=2\lambda(2k+n).
\end{gather*}
Here, $k$ and $n$ run over 0, 1, 2, \dots, and $c_{k,n}$ is some normalization coefficient. The determination of the latter, which is far more complicated than in the two-dimensional case, will be studied in detail in Section~\ref{section6}.

The action of $A^{\pm}$ and $Q^{\pm}$ on the wavefunctions can be easily calculated and is given by
\begin{gather}
 A^{+} \Psi_{k,n,0} =\frac{c_{k,n}}{c_{k,n+1}}\Psi_{k,n+1,0}, \label{eq:A+Psi} \\
 A^{-}\Psi_{k,n,0}=-\frac{4 \lambda k c_{k,n}}{c_{k-1,n+1}} \Psi_{k-1,n+1,0}, \label{eq:A-Psi} \\
 Q^{+}\Psi_{k,n,0} =\frac{c_{k,n}}{c_{k+1,n}} \Psi_{k+1,n,0}, \nonumber \\
 Q^{-}\Psi_{k,n,0}=\frac{8 \lambda^2 k(2n +2k +1)c_{k,n}}{c_{k-1,n}}\Psi_{k-1,n,0}-\frac{16g^2 k (k-1)
 c_{k,n}}{c_{k-2,n+2}}\Psi_{k-2,n+2,0}.\nonumber
\end{gather}

The algebra generated by $H$, $A^{\pm}$, and $Q^{\pm}$, and whose commutation relations are given in equations (\ref{eq:ha}), (\ref{eq:aa}), (\ref{eq:hq}), (\ref{eq:aqd}), (\ref{eq:eqm}), and (\ref{eq:qq}), is not a finite-dimensional Lie algebra. If the operators $A^{\pm}$ and $Q^{\pm}$ are useful for building the Hamiltonian wavefunctions, they do not lead to the hidden symmetry algebra. Furthermore, the wavefunctions $\Psi_{k,n,0}(z, \bar{z},x_3)$ with $n\ne0$ are self-orthogonal, which signals that $H$ is nondiagonalizable, so that some associated functions must be introduced to complete the basis and to get a resolution of identity. To try to solve these problems, it will prove convenient to introduce some additional ladder operators. This will be the purpose of Section~\ref{section3}.

\section{Construction of additional sets of ladder operators}\label{section3}

If we consider the subset $\{\Psi_{0,n,0} \,|\, n=0,1,2,\dots\}$ of wavefunctions, it is clear from equa\-tions~(\ref{eq:A+Psi}) and~(\ref{eq:A-Psi}) that we have a raising operator $A^{+}$, but no lowering one since $A^{-}$ annihilates~$\Psi_{0,n,0}$. This is actually another difference with the Hermitian case. The fact that the ladder operators are given by differential operators depending on more than one variable allows the existence of infinitely many zero modes.

Such a lowering operator is provided by
\begin{gather*}
 B^{-}=\partial_{\bar{z}}+ \frac{\lambda}{2} z - g x_3,
\end{gather*}
for which
\begin{gather*}
 B^{-}\Psi_{0,n,0} \propto \bar{z}^{n-1} {\rm e}^{-\frac{\lambda}{2}( z \bar{z} + x_3^2)+g \bar{z} x_3}
 \propto \Psi_{0,n-1,0}.
 \end{gather*}
This allows to consider another operators $B^{+}$, defined by
\begin{gather*}
 B^{+}=\partial_{\bar{z}} - \frac{\lambda}{2} z + g x_3.
\end{gather*}
With $H$, the pair of operators $B^{\pm}$ satisfy the commutation relations
\begin{gather}
 [H,B^{\pm}]=\pm 2 \lambda B^{\pm} \mp 2 g C^{\pm}, \label{eq:hb}
\end{gather}
where there appear two new operators
\begin{gather*}
 C^{\pm}=\partial_3 \pm g \bar{z} \mp \lambda x_3.
\end{gather*}
The other commutation relations are given by
\begin{gather}
 [H,C^{\pm}]=\mp 2 g A^{\pm} \pm 2\lambda C^{\pm}, \label{eq:hc} \\
 [A^{-},B^{+}]=[B^{-},A^{+}]=[C^{-},C^{+}]=-2\lambda, \label{eq:abcl} \\
 [B^{-},C^{+}]=[C^{-},B^{+}]=2g. \label{eq:abcq}
\end{gather}
From the three sets of operators $A^{\pm}$, $B^{\pm}$, and $C^{\pm}$, we can generate the operators $Q^{\pm}$, defined in (\ref{eq:q}), since
\begin{gather}
 Q^{\pm} = 2A^{\pm}B^{\pm} - (C^{\pm})^2. \label{eq:qbis}
\end{gather}
From equation (\ref{eq:qbis}) and the commutation relations (\ref{eq:aa}), (\ref{eq:abcl}), and (\ref{eq:abcq}), it follows that
\begin{gather*}
 [B^{\pm},Q^{\pm}]=[C^{\pm},Q^{\pm}]=0, \\
 [B^{\pm},Q^{\mp}]=\pm 4 \lambda B^{\mp} \pm 4 g C^{\mp}, \\
 [C^{\pm},Q^{\mp}]=\mp 4g A^{\mp} \mp 4 \lambda C^{\mp},
\end{gather*}
together with the set of relations (\ref{eq:aqd}), (\ref{eq:eqm}) and \eqref{eq:qq}.

In contrast with the algebra generated by $H$, $A^{\pm}$, and $Q^{\pm}$, the one generated by $H$, $A^{\pm}$, $B^{\pm}$,~$C^{\pm}$, and whose commutation relations are given by equations (\ref{eq:ha}), (\ref{eq:aa}), (\ref{eq:hb}), (\ref{eq:hc}), (\ref{eq:abcl}), and~(\ref{eq:abcq}), is a finite-dimensional Lie algebra. It has, however, a~rather complicated structure, so that some additional transformations have to be carried out, as shown in Section~\ref{section4}.

\section{Construction of the hidden symmetry algebra}\label{section4}

In order to get more insight into the structure of the hidden symmetry algebra of this non-Hermitian Hamiltonian, let us introduce a set of nine bilinear operators
\begin{gather}
 R=A^{+}A^{-},\qquad S=B^{+}B^{-},\qquad T=C^{+}C^{-}, \nonumber \\
 U=A^{+}B^{-}+B^{+}A^{-},\qquad V=A^{+}C^{-}+C^{+}A^{-},\qquad W=B^{+}C^{-}+C^{+}B^{-},
 \label{eq:R} \\
 X=A^{+}B^{-}-B^{+}A^{-},\qquad Y=A^{+}C^{-}-C^{+}A^{-},\qquad Z=B^{+}C^{-}-C^{+}B^{-}, \nonumber
\end{gather}
which have the following differential operator realizations
\begin{gather*}
 R=4 \partial_z^2 -\lambda^2 \bar{z}^2 ,\qquad S=\partial_{\bar{z}}^2 -\frac{\lambda^2}{4} z^2
 + \lambda g z x_3 -g^2 x_3^2,\\
 T=\partial_3^2 -g^2 \bar{z}^2 +2 \lambda g \bar{z} x_3 -\lambda^2 x_3^2 + \lambda, \nonumber \\
 U=4 \partial_z \partial_{\bar{z}} -\lambda^2 z \bar{z} +2 \lambda g \bar{z} x_3 + 2\lambda ,\qquad
 V= 4 \partial_z \partial_3 +2 \lambda g \bar{z}^2 -2 \lambda^2 \bar{z} x_3, \\
 W=2 \partial_{\bar{z}} \partial_3 + \lambda g z \bar{z} -\lambda^2 z x_3 -2 g^2 \bar{z} x_3
 +2 \lambda g x_3^2-2 g ,\qquad X=2(\lambda z-2 g x_3) \partial_z -2 \lambda \bar{z} \partial_{\bar{z}},\\
 Y=-4 (g \bar{z} -\lambda x_3) \partial_z -2 \lambda \bar{z} \partial_3,\qquad Z=-2( g \bar{z}-\lambda x_3)
 \partial_{\bar{z}} -(\lambda z -2 g x_3) \partial_z.
\end{gather*}
They satisfy the commutation relations{\samepage
\begin{gather}
 [R,S]=-2\lambda X,\qquad [R,T]=0,\qquad [R,U]=0,\qquad [R,V]=0, \nonumber\\
 [R,W]=-2\lambda Y,\qquad [R,X]=4\lambda R,\qquad [R,Y]=0,\qquad [R,Z]=-2\lambda V, \nonumber \\
 [S,T]=2g Z,\qquad [S,U]=0,\qquad [S,V]=-2\lambda Z -2g X ,\qquad [S,W]=0, \nonumber \\
 [S,X]=-4\lambda S,\qquad [S,Y]=-2\lambda W-2 g U,\qquad [S,Z]=-4g S, \nonumber \\
 [T,U]=-2g Y,\qquad [T,V]=2\lambda Y,\qquad [T,W]=2\lambda Z,\qquad [T,X]=-2g V, \nonumber \\
 [T,Y]=2\lambda V,\qquad [T,Z]=2\lambda W +4 g T,\qquad [U,V]=-2\lambda Y, \label{eq:RS}\\
 [U,W]=-2\lambda Z+2g X,\qquad [U,X]=0,\qquad [U,Y]=-2\lambda V -4 g R, \nonumber \\
 [U,Z]=-2\lambda W -2 g U,\qquad [V,W]=-2\lambda X + 2 g Y ,\qquad [V,X]=2\lambda V -4 g R, \nonumber \\
 [V,Y]=4\lambda R,\qquad [V,Z]=2\lambda (U-2T) +2 g V,\qquad [W,X]=-2\lambda W-2 g U, \nonumber \\
 [W,Y]=2\lambda ( U-2T) -2 g V,\qquad [W,Z]=4\lambda S, \qquad [X,Y]=-2\lambda Y, \nonumber \\
 [X,Z]=2\lambda Z-2g X,\qquad [Y,Z]=2\lambda X +2 g Y, \nonumber
\end{gather}
and therefore generate a nine-dimensional Lie algebra.}

These bilinear operators can be related to the Hamiltonian $H$ through the equation
\begin{gather*}
 H=-U-T.
\end{gather*}
This relation, which connects the Hamiltonian with the generators of the nine-dimensional Lie algebra, proves that the latter is a hidden symmetry algebra. Note, however, a distinction with respect to the Hermitian three-dimensional oscillator, for which $H$ is expressed in terms of three commuting components in involution, while here the components $U$ and $T$ have a nonvanishing commutator.

We may also point out the commutators of $H$ with the nine operators (\ref{eq:R}),
\begin{gather}
 [H,R]=0,\qquad [H,S]=2g Z,\qquad [H,T]=-2 g Y,\nonumber \\
 [H,U]=2g Y, \qquad [H,V]=0,\qquad [H,W]=-2g X, \label{eq:HR}\\
 [H,X]=2g V,\qquad [H,Y]=4 g R,\qquad [H,Z]=2g (U-2T), \nonumber
\end{gather}
which will be useful in further calculations.

\subsection[Connection with gl(3) and bosonic operators]{Connection with $\boldsymbol{{\mathfrak{gl}}(3)}$ and bosonic operators}\label{section4.1}

We can re-express the nine operators (\ref{eq:R}) in terms of ${\mathfrak{gl}}(3)$ generators $E_{ij}$, $i,j=1, 2, 3$, satisfying the commutation relations
\begin{gather*}
 [E_{ij},E_{kl}]=\delta_{j,k}E_{il}-\delta_{i,l}E_{kj}.
\end{gather*}
We indeed get the following relations
\begin{gather*}
 E_{11}=-\frac{1}{2\lambda}T+\frac{1}{2},\qquad E_{22}=\frac{\lambda}{2g^2}S + \frac{1}{2\lambda}T
 + \frac{1}{2g}W + \frac{1}{2}, \\
 E_{33}=-\frac{g^2}{2\lambda^3} R-\frac{\lambda}{2g^2}S-\frac{1}{2\lambda}T-\frac{1}{2\lambda}U
 -\frac{g}{2\lambda^2}V-\frac{1}{2g}W + \frac{1}{2}, \\
 E_{12}={\rm i}\left( \frac{1}{2\lambda} T + \frac{1}{4g} W-\frac{1}{4g} Z\right),\qquad
 E_{21}={\rm i}\left(\frac{1}{2\lambda} T + \frac{1}{4g} W+\frac{1}{4g} Z\right), \\
 E_{13}=-\frac{1}{2\lambda} T- \frac{g}{4\lambda^2} V - \frac{1}{4g} W + \frac{g}{4\lambda^2}Y
 + \frac{1}{4g} Z, \\
 E_{31}=-\frac{1}{2\lambda} T- \frac{g}{4\lambda^2} V - \frac{1}{4g} W - \frac{g}{4\lambda^2}Y
 - \frac{1}{4g} Z, \\
 E_{23}={\rm i}\left( \frac{\lambda}{2g^2}S + \frac{1}{2\lambda}T + \frac{1}{4\lambda}U
 + \frac{g}{4\lambda^2} V + \frac{1}{2g} W - \frac{1}{4\lambda}X - \frac{g}{4\lambda^2} Y\right),\\
 E_{32}={\rm i}\left( \frac{\lambda}{2g^2}S + \frac{1}{2\lambda}T + \frac{1}{4\lambda}U
 + \frac{g}{4\lambda^2} V + \frac{1}{2g} W + \frac{1}{4\lambda}X + \frac{g}{4\lambda^2} Y\right).
\end{gather*}

The ${\mathfrak{gl}}(3)$ linear Casimir operator corresponds to
\begin{align}
 E_{11}+E_{22} +E_{33} &=-\frac{1}{2\lambda}\left( T+ U + \frac{g^2}{\lambda^2}R + \frac{g}{\lambda}V
 -3\lambda\right) \nonumber \\
 &= \frac{1}{2\lambda} \left( H -\frac{g^2}{\lambda^2}R-\frac{g}{\lambda} V +3\lambda \right),
 \label{eq:casimir}
\end{align}
so that it is a linear combination of the three commuting operators $H$, $R$, and $V$, up to some additive constant.

Rewriting the generators $E_{ij}$ in terms of $A^{\pm}$, $B^{\pm}$, and $C^{\pm}$, as done in Appendix~\ref{appendixA}, allows to reveal some underlying structure in terms of bosonic operators $a^{\pm}_i$, $i=1,2,3$, satisfying the well-known commutation relations
\begin{gather*}
 [a_i^{-},a_j^{+}]=\delta_{i,j},\qquad [a_i^{\pm},a_j^{\pm}]=0.
\end{gather*}
On using the transformation
\begin{gather}
\begin{split}
& a_1^{\pm}=\frac{\rm i}{\sqrt{2\lambda}}C^{\pm},\qquad a_2^{\pm}=\frac{1}{\sqrt{2\lambda}}
 \left(C^{\pm}+ \frac{\lambda}{g}B^{\pm} \right), \\
& a_3^{\pm}=\frac{\rm i}{\sqrt{2\lambda}}\left(C^{\pm} + \frac{\lambda}{g}B^{\pm}+ \frac{g}{\lambda}
 A^{\pm}\right),
\end{split} \label{eq:boson}
\end{gather}
it is indeed possible to rewrite $E_{ij}$ as
\begin{gather*}
 E_{ij}= \frac{1}{2}\{a_i^{+},a_j^{-} \} = a_i^+ a_j^- + \frac{1}{2}\delta_{ij}.
\end{gather*}
From the inverse transformation of (\ref{eq:boson}),
\begin{gather*}
 A^{\pm}=-\frac{\lambda}{g}\sqrt{2\lambda}(a_2^{\pm}+{\rm i} a_3^{\pm}),\qquad B^{\pm}=
 \frac{g}{\lambda}\sqrt{2\lambda}(a_2^{\pm}+{\rm i}a_1^{\pm}),\qquad C^{\pm}
 =-{\rm i}\sqrt{2\lambda}a_1^{\pm},
\end{gather*}
we can also express $Q^{\pm}$ and $H$ in terms of the bosonic operators,
\begin{gather*}
 Q^{\pm}=2\lambda \big[ (a_1^{\pm})^2 -2 (a_2^{\pm})^2 -2 {\rm i} a_1^{\pm} a_2^{\pm}
 + 2 a_1^{\pm} a_3^{\pm} -2 {\rm i} a_2^{\pm} a_3^{\pm} \big], \\
 H=2\lambda \big[ a_1^{+} a_1^{-} + 2 a_2^{+} a_2^{-} + {\rm i} (a_1^{+} a_2^- +a_2^{+}a_1^{-}) - (a_1^{+}
 a_3^{-} +a_3^{+} a_1^{-}) + {\rm i} ( a_2^{+}a_3^{-} +a_3^{+}a_2^{-}) \big].
\end{gather*}

The ${\mathfrak{gl}}(3)$ hidden symmetry algebra can be embedded into an ${\mathfrak{sp}}(6)$ algebra by considering the additional generators \cite{mosh}
\begin{gather*}
 D_{ij}^{+}=\frac{1}{2}\{a_i^{+},a_j^{+}\},\qquad D_{ij}^{-}=\frac{1}{2}\{a_i^{-},a_j^{-}\}.
\end{gather*}
Together with the bosonic operators, the operators $E_{ij}$, $D^+_{ij}$, and $D^-_{ij}$ then make rise to an ${\mathfrak{osp}}(1/6)$ superalgebra~\cite{frappat} (see \cite{marquette20} for more details).

Finally, we may also point out the nonstandard differential operator realization of the bosonic operators,
\begin{gather*}
 a_1^{\pm}=\frac{\rm i}{\sqrt{2\lambda}} (\partial_3 \pm g \bar{z} \mp \lambda x_3),\quad
 a_2^{\pm} =\frac{1}{\sqrt{2\lambda}} \left(\frac{\lambda}{g} \partial_{\bar{z}} + \partial_3 \mp
 \frac{\lambda^2}{2g} z \pm g \bar{z}\right), \\
 a_3^{\pm}=\frac{\rm i}{\sqrt{2\lambda}}\left(2 \frac{g}{\lambda} \partial_z + \frac{\lambda}{g}
 \partial_{\bar{z}} + \partial_3 \mp \frac{\lambda^2}{2g} z \right).
\end{gather*}

This completes the description of the hidden symmetry algebra. As compared with the Hermitian case, the analysis carried out here has shown that the Hamiltonian connects in a~different way with the algebra Casimir operator and that a nonstandard realization of bosonic operators makes its appearance. Further progress on these ideas might provide a way to classify some classes of nonseparable and nondiagonalizable models.

\subsection{Superintegrability and cubic algebra}\label{section4.2}

In \cite{barda}, four independent operators $R_0$, $R_1$, $R_2$, and $R_3$ commuting with $H$ were identified, the first two being mutually commuting. Here we plan to relate such a superintegrability property with the hidden symmetry algebra.

On expressing the operators of \cite{barda} in terms of the bilinear operators (\ref{eq:R}), we get
\begin{gather*}
 R_{0}=A^{+}A^{-} =R, \\
 R_1=\frac{1}{8g} \left\{ \frac{1}{2} [Q^{+},Q^{-}] + 4\lambda H + 12\lambda^2\right\}=\frac{1}{2} V, \\
 R_2=\frac{1}{8\lambda}[A^{+}A^{-},Q^{+}Q^{-}] = -R( X-2\lambda ) + \frac{1}{4} \{V,Y\}, \\
 R_3=Q^{+}(A^{-})^2 = R( U -X + 4 \lambda) -\frac{1}{4}\big( V^2 + Y^2 -\{V,Y\} \big).
\end{gather*}
From the commutation relations~(\ref{eq:HR}), it follows that
\begin{gather}
 [H,R_0]=[H,R_1]=[H,R_2]=[H,R_3]=0, \label{eq:HR0}
\end{gather}
while equation (\ref{eq:RS}) leads to
\begin{gather}
 [R_0,R_1]=0,\qquad [R_0,R_2]=-4\lambda R_0^2,\qquad [R_1,R_2]= 2 g R_0^2, \label{eq:R0R1}
\end{gather}
and
\begin{gather}
\begin{split}
& [R_0,R_3]-4\lambda R_0^2,\qquad [R_1,R_3]=2 g R_0^2, \\
& [R_2,R_3]=8gR_1R_0^2 + 4\lambda (R_3-R_2)R_0 + 8\lambda R_1^2R_0.
\end{split} \label{eq:R0R3}
\end{gather}
Equations (\ref{eq:HR0}) and(\ref{eq:R0R1}) agree with some results derived in \cite{barda}, while equation (\ref{eq:R0R3}) is new and shows that the integrals of motion generate a cubic algebra.

Note that the superintegrability property of the present model contrasts with what was obtained in \cite{marquette20} for the two-dimensional pseudo-Hermitian oscillator, which was proved to be only integrable. This points out that in the context of non-Hermitian Hamiltonians, the number of integrals of motion may be affected by the structure of the problem.

\section{Nondiagonalizability and construction of associated functions}\label{section5}

As pointed out above, the Hamiltonian $H$ being nondiagonalizable, the excited wavefunctions $\Psi_{k,n,0}$ with $n\ne0$ have to be accompanied with some associated functions $\Psi_{k,n,m}$, $m=1,2,\dots, p_n-1$, completing the Jordan blocks. By definition, these functions obey the relation
\begin{gather}
 (H-E_{k,n}) \Psi_{k,n,m} = \Psi_{k,n,m-1}, \qquad m=1, 2, \dots, p_n-1. \label{eq:associate}
\end{gather}
It is the purpose of the present section to determine them and to establish that the dimension of the Jordan blocks is $p_n = 2n+1$. We plan to rely on the new sets of ladder operators of Section~\ref{section3} to provide an algebraic construction of these associated functions. Furthermore, we will also build a subset of them in terms of multivariate polynomials. Finally, we will determine the action of the ladder operators and of the ${\mathfrak{gl}}(3)$ Casimir operator on the associated functions.

\subsection{Algebraic construction of associated functions}\label{section5.1}

Let us start by noting some equivalences among polynomials of the operators acting on the ground state,
\begin{gather*}
 (H-2\lambda)B^{+} \Psi_0 = -2 g C^{+} \Psi_0, \qquad (H-2\lambda)^2 B^{+}\Psi_0 = 4 g^2 A^{+} \Psi_0
 \propto \Psi_{0,1,0}.
\end{gather*}
Hence $(H-2\lambda)B^{+} \Psi_0 \propto \Psi_{0,1,1}$ and $B^{+}\Psi_0 \propto \Psi_{0,1,2}$.
We would like to extend these results by showing that
\begin{gather*}
 (H-2\lambda n)^{2n}(B^{+})^n \Psi_0 \propto (A^{+})^n \Psi_0 \propto \Psi_{0,n,0},
\end{gather*}
and more generally that
\begin{gather}
 (H-2\lambda n)^{2p} (B^{+})^n \Psi_0 \nonumber\\
 \qquad{} =(B^{+})^{n-2p} \sum_{q=0}^{p} a_q^{(n,p)}(A^{+}B^{+})^q
 (C^{+})^{2p-2q} \Psi_0,\qquad p=0,1,\dots,n, \label{eq:aqp} \\
 (H-2\lambda n)^{2p+1} (B^{+})^n \Psi_0 \nonumber\\
 \qquad {}= (B^{+})^{n-2p-1} \sum_{q=0}^{p} b_q^{(n,p)} (A^{+}
 B^{+})^q (C^{+})^{2p+1-2q} \Psi_0,\qquad p=0,1,\dots,n-1, \label{eq:bqp}
\end{gather}
for some coefficients $a^{(n,p)}_q$ and $b^{(n,p)}_q$ to be determined.

Let us first point out some auxiliary results,
\begin{gather*}
 (H-2\lambda n) (A^{+})^p= (A^{+})^p [ H-2\lambda (n-p) ], \nonumber \\
 (H-2\lambda n) (B^{+})^p= (B^{+})^p [ H-2\lambda (n-p) ]- 2 p q (B^{+})^{p-1} C^{+}, \\
 (H-2\lambda n) (C^{+})^p= (C^{+})^p [ H-2\lambda (n-p) ] -2 p g A^{+} (C^{+})^{p-1}, \nonumber
\end{gather*}

which directly follow from the commutation relations established in Sections~\ref{section2} and~\ref{section3}.

By acting on (\ref{eq:aqp}) with $(H - 2 \lambda n)$ and identifying the result with equation (\ref{eq:bqp}), we get the coefficients $b_q^{(n,p)}$ in terms of $a_r^{(n,p)}$,
\begin{gather*}
 b_0^{(n,p)}=-2 g (n-2p) a_0^{(n,p)}, \\
 b_q^{(n,p)}=-2g \big[ (2p-2q +2) a_{q-1}^{(n,p)} + ( n-2p +q) a_q^{(n,p)} \big],\qquad q=1,2,\dots,p.
\end{gather*}
Alternatively, by acting on (\ref{eq:bqp}) with $(H - 2 \lambda n)$ and identifying the result with equation (\ref{eq:aqp}), where $p$ is replaced by $p+1$, we find the coefficients $a_q^{(n,p+1)}$ in terms of $b_r^{(n,p)}$,
\begin{gather*}
 a_0^{(n,p+1)}=-2g (n-2p-1) b_0^{(n,p)}, \\
 a_q^{(n,p+1)}=-2g \big[ (2p +3 -2q ) b_{q-1}^{(n,p)} + (n-2p +q -1)b_{q}^{(n,p)}\big], \quad q=1,2,\dots,p, \\
 a_{p+1}^{(n,p+1)}=-2 g b_p^{(n,p)}.
\end{gather*}
On eliminating $b_{q}^{(n,p)}$ or $a_{q}^{(n,p)}$ between the two sets of relations, we obtain recursion relations for $a_{q}^{(n,p)}$ or $b_{q}^{(n,p)}$,
\begin{gather*}
 a_0^{(n,p+1)}=4 g^2 (n-2p -1)( n-2p) a_0^{(n,p)}, \\
 a_1^{(n,p+1)}=4 g^2 (n-2p) \big[ (4p+1) a_0^{(n,p)} + ( n-2p+1) a_1^{(n,p)}\big], \\
 a_q^{(n,p+1)}=4 g^2 \big[ (2p-2q+3)(2p-2q +4) a_{q-2}^{(n,p)} + (n-2p+q-1)(4p-4q +5)a_{q-1}^{(n,p)} \\
\hphantom{a_q^{(n,p+1)}=}{} + (n-2p +q-1)(n-2p+q)a_q^{(n,p)}\big],\quad q=2,3,\dots,p, \\
 a_{p+1}^{(n,p+1)} = 4g^2 \big[ 2 a_{p-1}^{(n,p)} + (n-p) a_p^{(n,p)}\big],
\end{gather*}
and
\begin{gather*}
 b_0^{(n,p+1)}=4 g^2 (n-2p -2)( n-2p-1) b_0^{(n,p)}, \\
 b_1^{(n,p+1)}=4 g^2 (n-2p-1) \big[ (4p+3) b_0^{(n,p)} + ( n-2p) b_1^{(n,p)}\big], \\
 b_q^{(n,p+1)}=4 g^2 \big[ (2p-2q+4)(2p-2q +5) b_{q-2}^{(n,p)} + (n-2p+q-2)(4p-4q +7)b_{q-1}^{(n,p)} \\
\hphantom{b_q^{(n,p+1)}=}{} + (n-2p +q-2)(n-2p+q-1)b_q^{(n,p)}\big],\quad q=2,3,\dots,p, \\
 b_{p+1}^{(n,p+1)} = 12g^2 \big[ 2 b_{p-1}^{(n,p)} + (n-p-1) b_p^{(n,p)} \big],
\end{gather*}
respectively.

The solutions of the recursion relations are given by
\begin{gather}
 a_q^{(n,p)}= (2g)^{2p} \frac{n!}{(n-2p+q)!}\frac{p!}{q!(p-q)!} \frac{(2p-1)!!}{(2p-2q-1)!!} \label{eq:a}
\end{gather}
and
\begin{gather}
 b_q^{(n,p)}= (-2g)^{2p+1} \frac{n!}{(n-2p+q-1)!}\frac{p!}{q!(p-q)!} \frac{(2p+1)!!}{(2p-2q+1)!!}.
 \label{eq:b}
\end{gather}
At this stage, it is worth noting that in equations (\ref{eq:aqp}) and (\ref{eq:bqp}), the summations over $q$ do not really go from $0$ to $p$ because the exponents of the operators have to be nonnegative, hence $q$ may not be smaller than $2p-n$ and $2p+1-n$, respectively. In fact, the summations run from max$(0,2p-n)$ to $p$ or max$(0,2p+1-n)$ to $p$. This property is accounted for by the presence of $(n-2p+q)!$ or $( n-2p +q-1)!$ in the denominator for $a_q^{(n,p)}$ or $b_q^{(n,p)}$. We may indeed interpret these factorials as $\Gamma(n-2p+q+1)$ or $\Gamma(n-2p+q)$, which become infinite for $q \leq 2p-n-1$ or $q \leq 2p-n$ and produce the vanishing of the corresponding $a_q^{(n,p)}$ or $b_q^{(n,p)}$.

More generally, on applying $(Q^+)^{k}$ to both sides of equations (\ref{eq:aqp}) and (\ref{eq:bqp}) and taking into account that $[H,Q^+] = 4\lambda Q^+$ and $[Q^+,A^+] = [Q^+,B^+] = [Q^+,C^+] = 0$, we obtain
\begin{gather*}
[H-2\lambda (n+2k)]^{2p} (B^{+})^n (Q^+)^k \Psi_0\\
 \qquad{} =(B^{+})^{n-2p} \sum_{q=0}^{p} a_q^{(n,p)}
 (A^{+}B^{+})^q (C^{+})^{2p-2q} (Q^+)^k \Psi_0, \qquad p=0,1,\dots,n,
\end{gather*}
and
\begin{gather*}
 [H-2\lambda (n+2k)]^{2p+1} (B^{+})^n (Q^+)^k\Psi_0 \\
 \qquad {} = (B^{+})^{n-2p-1} \sum_{q=0}^{p} b_q^{(n,p)}
 (A^{+} B^{+})^q (C^{+})^{2p+1-2q} (Q^+)^k \Psi_0, \qquad p=0,1,\dots,n-1,
\end{gather*}
so that in particular
\begin{gather}
 [H-2\lambda(n+2k)]^{2n} (B^+)^n Q^+)^k \Psi_0 = a^{(n,n)}_n (A^+)^n (Q^+)^k \Psi_0 \propto
 \Psi_{k,n,0}. \label{eq:wf-bis}
\end{gather}
with
\begin{gather*}
 a^{(n,n)}_n = (2g)^{2n} n!(2n-1)!!.
\end{gather*}

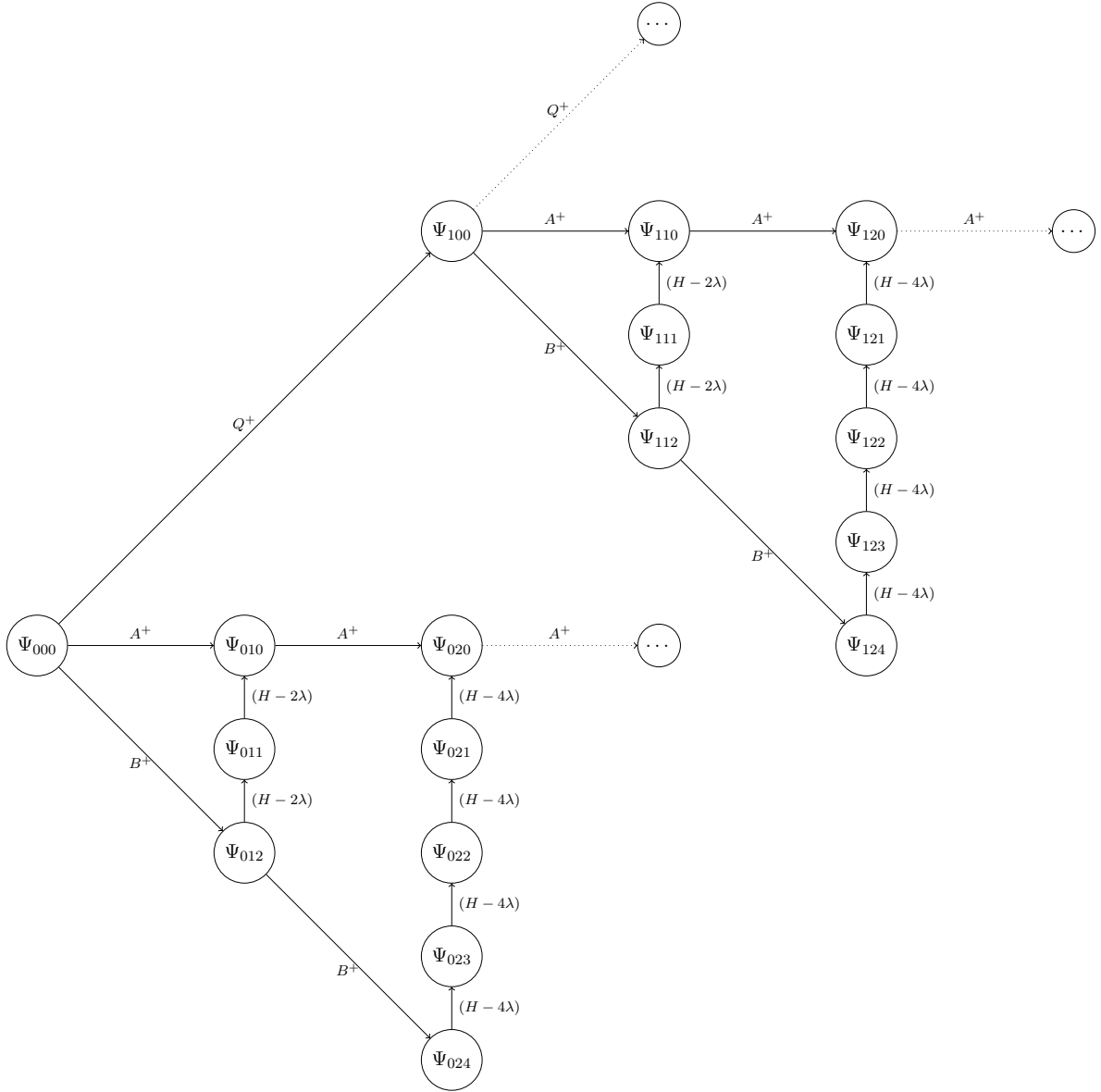
\begin{figure}[t]\centering
\tikzset{
 net node/.style = {draw, circle, minimum size=8mm},
 net edge/.style = {->, >=triangle 45}, 
 net cut/.style = {shorten >=-10mm, shorten <=-10mm, rounded corners=10mm, color=red},
 net cross/.style = {sloped, allow upside down, pos=.3},
}

\resizebox{.99\textwidth}{!}{

\begin{tikzpicture}

 \newcommand{\edge}[5][]{\draw[net edge, #1] (#3) -- coordinate[net cross, name=#2] node[pos=.7, auto]{#5} (#4);}

\tikzstyle{label}=[midway,font=\scriptsize],
 \node[net node ] (000) at (0,8) {$\Psi_{000}$};
 \node[net node ] (010) at (4,8) {$\Psi_{010}$};
 \node[net node ] (011) at (4,6) {$\Psi_{011}$};
 \node[net node ] (012) at (4,4) {$\Psi_{012}$};
		
		\node[net node] (020) at (8,8) {$\Psi_{020}$};
		\node[net node] (021) at (8,6) {$\Psi_{021}$};
		\node[net node] (022) at (8,4) {$\Psi_{022}$};
		\node[net node] (023) at (8,2) {$\Psi_{023}$};
		\node[net node] (024) at (8,0) {$\Psi_{024}$};

					\node[net node] (100) at (8,16) {$\Psi_{100}$};
					\node[net node] (110) at (12,16) {$\Psi_{110}$};
					\node[net node] (111) at (12,14) {$\Psi_{111}$};
					\node[net node] (112) at (12,12) {$\Psi_{112}$};

				\node[net node] (120) at (16,16) {$\Psi_{120}$};
				\node[net node] (121) at (16,14) {$\Psi_{121}$};
				\node[net node] (122) at (16,12) {$\Psi_{122}$};
				\node[net node] (123) at (16,10) {$\Psi_{123}$};
				\node[net node] (124) at (16,8) {$\Psi_{124}$};

		\node[net node] (ex1) at (12,20) {$\dots$};
		\node[net node] (ex2) at (12,8) {$\dots$};
		\node[net node] (ex3) at (20,16) {$\dots$};

 \draw[->] (000)--(010) node [label,above] {$A^{+}$};
 \draw[->] (000)--(012) node [label,below] {$B^{+}$};
 \draw[->] (012)--(011) node [label,right] {$(H-2\lambda)$};
 \draw[->] (011)--(010) node [label,right] {$(H-2\lambda)$};

	 \draw[->] (010)--(020) node [label,above] {$A^{+}$};
 \draw[->] (012)--(024) node [label,below] {$B^{+}$};
 \draw[->] (024)--(023) node [label,right] {$(H-4\lambda)$};
 \draw[->] (023)--(022) node [label,right] {$(H-4\lambda)$};
		 \draw[->] (022)--(021) node [label,right] {$(H-4\lambda)$};
 \draw[->] (021)--(020) node [label,right] {$(H-4\lambda)$};

		\draw[->] (000)--(100) node [label,above] {$Q^{+}$};
		
		\draw[->] (100)--(110) node [label,above] {$A^{+}$};
 \draw[->] (100)--(112) node [label,below] {$B^{+}$};
 \draw[->] (112)--(111) node [label,right] {$(H-2\lambda)$};
 \draw[->] (111)--(110) node [label,right] {$(H-2\lambda)$};

		\draw[->] (110)--(120) node [label,above] {$A^{+}$};
 \draw[->] (112)--(124) node [label,below] {$B^{+}$};
 \draw[->] (124)--(123) node [label,right] {$(H-4\lambda)$};
 \draw[->] (123)--(122) node [label,right] {$(H-4\lambda)$};
		 \draw[->] (122)--(121) node [label,right] {$(H-4\lambda)$};
 \draw[->] (121)--(120) node [label,right] {$(H-4\lambda)$};
		
				\draw[->,dotted] (100)--(ex1) node [label,above] {$Q^{+}$};
			\draw[->,dotted] (020)--(ex2) node [label,above] {$A^{+}$};
		 \draw[->,dotted] (120)--(ex3) node [label,above] {$A^{+}$};
		
\end{tikzpicture}
}

\caption{Construction of associated states by using ladder operators.}\label{fig1}
\end{figure}

This establishes that the dimension of the Jordan blocks is $p_n = 2n+1$ and that the associated functions can be obtained from the ground state in the form
\begin{gather}
 \Psi_{k,n,m} = d_{k,n} [ H- 2 \lambda (n+2k))]^{2n-m} (B^{+})^n (Q^{+})^k \Psi_0,\qquad m=1,2,\dots,2n.
 \label{eq:associate-1}
\end{gather}
Their detailed expressions in terms of the ladder operators are given by
\begin{gather}
 \Psi_{k,n,m}= \begin{cases}
\displaystyle d_{k,n}(B^{+})^{2\mu-n} \sum\limits_{q=\max(0,n-2\mu)}^{n-\mu} a_q^{(n,n-\mu)} (A^{+} B^{+})^{q}
 & \\
\displaystyle \qquad {}\times (C^{+})^{2n-2\mu-2q} (Q^{+})^k \Psi_0 & \text{if $m=2\mu$}, \\
\displaystyle d_{k,n}(B^{+})^{2\mu+1-n} \sum\limits_{q=\max(0,n-2\mu-1)}^{n-\mu-1} b_q^{(n,n-\mu-1)} (A^{+}
 B^{+})^{q} & \\
 \displaystyle\qquad {}\times (C^{+})^{2n-2\mu-2q-1} (Q^{+})^k \Psi_0 &\text{if $m=2\mu+1$},
\end{cases} \label{eq:associate-2}
\end{gather}
where the coefficients are expressed in (\ref{eq:a}) and (\ref{eq:b}), and p has been replaced by $n-\mu$ or $n-\mu-1$, respectively. The additional factor $d_{k,n}$ is a normalization coefficient, whose calculation will be discussed in Section~\ref{section6}. The construction of the associated functions $\Psi_{k,n,m}$ is displayed in Figure~\ref{fig1}.

At this stage, it is worth observing that, in accordance with equations (\ref{eq:associate}) and (\ref{eq:wf-bis}), the wavefunctions (\ref{eq:wf}) can be alternatively expressed in the form (\ref{eq:associate-1}) or (\ref{eq:associate-2}), with $m$ set equal to zero. It follows that the normalization coefficients $c_{k,n}$ and $d_{k,n}$ of equations (\ref{eq:wf}) and (\ref{eq:associate-1}), respectively, are connected by the relation
\begin{gather}
 c_{k,n} = d_{k,n} a^{(n,n)}_n = d_{k,n} (2g)^{2n} n! (2n-1)!!. \label{eq:c-d}
\end{gather}

\subsection{Some associated functions in terms of multivariate polynomials}\label{section5.2}

After presenting an algebraic description of the associated functions, we will now use an alternative approach wherein some of them will be written in terms of polynomials in the variables~$z$,~$\bar{z}$, and~$x_3$.

To start with, it is useful to introduce a new set of variables,
\begin{gather*}
 u=\bar{z},\qquad v=-\lambda z +2 g x_3 ,\qquad w=g \bar{z} -\lambda x_3,
\end{gather*}
which are directly related to the action of the raising operators $A^+$, $B^+$, and $C^+$ on $\Psi_0$,
\begin{gather*}
 A^+ \Psi_0 = -2\lambda u \Psi_0, \qquad B^+ \Psi_0 = v \Psi_0, \qquad C^+ \Psi_0 = 2w \Psi_0.
\end{gather*}
The Jacobian of the transformation is
\begin{gather*}
\frac{\partial(u,v,w)}{\partial(z,\bar{z},x_3)}= \left|
 \begin{matrix}
 0 & 1 & 0 \\
 -\lambda & 0 & 2g \\
 0 & g & -\lambda
 \end{matrix}
\right|=-\lambda^2
\end{gather*}
and the inverse transformation writes
\begin{gather*}
z=\frac{1}{\lambda^2}(2g^2 u -\lambda v -2g w),\qquad \bar{z}=u,\qquad x_3=\frac{1}{\lambda}(g u - w).
\end{gather*}

Furthermore, we may introduce a new operator $D_p$, defined by
\begin{gather*}
 (H-2\lambda p)\Psi_0 =\Psi_0 D_p,
\end{gather*}
and whose explicit expression in terms of the new variables is
\begin{gather*}
 D_p=4\lambda \partial^2_{uv} -4 g^2 \partial^2_{v} +8 \lambda g \partial^2_{vw} -\lambda^2 \partial^2_{w}
 +2 \lambda( u \partial_u + v \partial_v + w \partial_w -p) - 4 g w \partial_v + 2 \lambda g u \partial_w.
\end{gather*}
Its action on the monomials $u^i$, $v^{i}$, and $w^{i}$ is given by
\begin{gather*}
 D_p u^i =u^i D_p + 2\lambda i u^{i-1} (2 \partial_v +u ), \\
 D_p v^i =v^i D_p + i v^{i-1} \big( 4 \lambda \partial_u - 8 g^2 \partial_v + 8 \lambda g \partial_w +2 \lambda v
 - 4 g w \big) - 4 i (i-1) g^2 v^{i-2}, \\
 D_p w^i=w^i D_p +2 \lambda i w^{i-1} ( 4g \partial_v -\lambda \partial_w + g u + w ) - i (i-1)\lambda^2
 w^{i-2}.
\end{gather*}

Let us consider the set of functions belonging to a Jordan block with $k=0$ and any $n$ (or, in other words, those belonging to the lower lattice in Figure~\ref{fig1}),
\begin{gather*}
 \Psi_{0,n,2n-p}=d_{0,n}(H-E_{0n})^p v^n \Psi_0,\qquad p=0,1,\dots,2n,
\end{gather*}
and introduce the notation
\begin{gather*}
 v_i^{(n)}=\frac{\Gamma(n+1)}{\Gamma(n-i+1)}v^{n-i},\qquad i=0,1,\dots,n,
\end{gather*}
so that $v_0^{(n)}=v^n$, $v_i^{(n)}=n(n-1)\cdots(n-i+1)v^{n-i}$, with $i=1,2,\dots,n$, and
\begin{gather*}
 D_n v_i^{(n)}=-2\lambda i v_i^{(n)} -4 g v_{i+1}^{(n)} w - 4 g^2 v_{i+2}^{(n)}.
\end{gather*}
Then we obtain
\begin{gather*}
 \Psi_{0,n,2n-p}=d_{0,n} \Psi_0 \sum_{q=[\frac{p+1}{2}]}^{2p} v_q^{(n)} f_q^{(n,p)}(u,w),
\end{gather*}
where
\begin{gather*}
 f_q^{(n,p)}(u,w)=\sum_{r,s} \alpha_{r,s}^{(n,p,q)} u^r w^s
\end{gather*}
is a polynomial of total degree $2p-q$ and total parity $(-1)^q$ in the cubic variable $u$ and the linear one $w$. This means that $3 r+s \leq 2p-q$ and $(-1)^{3r+s}=(-1)^q$. The coefficients $\alpha^{(n,p,q)}_{r,s}$ satisfy the recursion relation
\begin{gather*}
 \alpha_{r,s}^{(n,p+1,q)} =\lambda \big[ 2(r+s-q) \alpha_{r,s}^{(n,p,q)} + 2 q(s+1) \alpha_{r-1,s+1}
 ^{(n,p,q)}-\lambda (s+1)(s+2) \alpha_{r,s+2}^{(n,p,q)}\big] \\
\hphantom{\alpha_{r,s}^{(n,p+1,q)} =}{} + 4 \big[ {-}g \alpha_{r,s-1}^{(n,p,q-1)} + 2 \lambda g (s+1) \alpha_{r,s+1}^{(n,p,q-1)}+\lambda (r+1)
 \alpha_{r+1,s}^{(n,p,q-1)}\big] -4 g^2 \alpha_{r,s}^{(n,p,q-2)},
\end{gather*}
which can be solved for low values of $p$ and $q$. Some examples of polynomials $f^{(n,p)}_q(u,w)$ are presented in Appendix~\ref{appendixB}.

\subsection[Action of the ladder operators and of the gl(3) Casimir operator on Psi\_\{k,n,m\}]{Action of the ladder operators and of the $\boldsymbol{{\mathfrak{gl}}(3)}$ Casimir operator\\ on $\boldsymbol{\Psi_{k,n,m}}$}\label{section5.3}

The structure of the functions $\Psi_{k,n,m}$ is complicated, but using our algebraic description in terms of the ladder operators $A^{\pm}$, $B^{\pm}$, and $C^{\pm}$ allows us to establish explicit formulas for the action of these operators.

For $A^{+}$, we get
\begin{gather*}
 A^{+}\psi_{k,n,m}=\alpha_0^{(k,n,m)}\psi_{k,n+1,m} + \alpha_1^{(k,n,m)}\psi_{k+1,n-1,m-2},
\end{gather*}
where the coefficients $\alpha_0^{(k,n,m)}$ and $\alpha_1^{(k,n,m)}$ are given by
\begin{gather}
 \alpha_0^{(k,n,m)}=\frac{d_{k,n}}{d_{k,n+1}}\frac{1}{4g^2(n+1)(2n+1)}, \label{eq:alpha0} \\
 \alpha_1^{(k,n,m)}=\begin{cases}
 \frac{d_{k,n}}{d_{k+1,n-1}}\frac{n}{2n+1} & \text{if $m \geq 2$}, \\
 0 & \text{if $m=0$ or 1} \label{eq:alpha1},
\end{cases}
\end{gather}
in terms of the normalization coefficients, whose calculation will be discussed in Section~\ref{section6}.

This result can be demonstrated by direct calculations. Let us provide some details on the proof.

{\it Case 1}: $m=2\mu$. Assuming $n \ge 2\mu$, we get
\begin{gather*}
 A^{+}\Psi_{k,n,2\mu}= d_{k,n}(B^{+})^{2\mu-n-1} \sum_{q=n-2\mu}^{n-\mu} a_q^{(n,n-\mu)}
 (A^{+} B^{+})^{q+1} (C^{+})^{2n-2\mu -2q} (Q^{+})^k \Psi_0 \\
 =\alpha_0^{(k,n,2\mu)} d_{k,n+1} (B^{+})^{2\mu -n -1} \sum_{q=n-2\mu +1}^{n-\mu +1}
 a_q^{(n+1,n-\mu+1)} (A^{+}B^{+})^q (C^{+})^{2n-2\mu -2q+2} (Q^{+})^k \Psi_0 \\
 \quad {}+ \alpha_1^{(k,n,2\mu)} d_{k+1,n-1} (B^{+})^{2\mu -n -1} \Biggl[ 2 \sum_{q=n-2\mu +2}
 ^{n-\mu+1} a_{q-1}^{(n-1,n-\mu)}(A^{+} B^{+})^q (C^{+})^{2n-2\mu -2q +2} \\
 \quad {}-\sum_{q=n-2\mu+1}^{n-\mu} a_q^{(n-1,n-\mu)} (A^{+} B^{+})^q (C^{+})^{2n-2\mu-2q+2}
 \Biggr] (Q^{+})^k \Psi_0.
\end{gather*}
By equating both expressions, we obtain the following constraints to be satisfied:
\begin{gather*}
 d_{k,n} a_{n-2\mu}^{(n,n-\mu)} =\alpha_0^{(k,n,2\mu)} d_{k,n+1} a_{n-2\mu+1}^{(n+1,n-\mu+1)}
 - \alpha_1^{(k,n,2\mu)} d_{k+1,n-1} a_{n-2\mu+1}^{(n-1,n-\mu)} \nonumber \\
\hphantom{d_{k,n} a_{n-2\mu}^{(n,n-\mu)} =}{} \text{if $q=n-2\mu+1$}, \\
 d_{k,n} a_{q-1}^{(n,n-\mu)} =\alpha_0^{(k,n,2\mu)} d_{k,n+1} a_q^{(n+1,n-\mu+1)} + \alpha_1^{(k,n,2\mu)} d_{k+1,n-1} \big( 2 a_{q-1}^{(n-1,n-\mu)}-a_q^{(n-1,n-\mu)}\big) \nonumber \\
\hphantom{d_{k,n} a_{q-1}^{(n,n-\mu)} =}{} \text{if $n-2\mu+2 \le q \le n-\mu$}, \\
 d_{k,n} a_{n-\mu}^{(n,n-\mu)} =\alpha_0^{(k,n)} d_{k,n+1} a_{n-\mu+1}^{(n+1,n-\mu+1)}
 +2 \alpha_1^{(k,n,2\mu)}d_{k+1,n-1} a_{n-\mu}^{(n-1,n-\mu)} \nonumber \\
\hphantom{d_{k,n} a_{n-\mu}^{(n,n-\mu)}}{} \text{if $q=n-\mu+1$}.
\end{gather*}
These are fulfilled due to the definitions of $\alpha_0^{(k,n,m)}$, $\alpha_1^{(k,n,m)}$, and $a_q^{(n,n-\mu)}$. Whenever $n<2\mu$, a~similar procedure leads to the proof of equations (\ref{eq:alpha0}) and (\ref{eq:alpha1}).

{\it Case 2}: $m=2\mu+1$.
Assuming $n\ge 2\mu+1$, we get
\begin{gather*}
 A^{+}\Psi_{k,n,2\mu+1}=d_{k,n}(B^{+})^{2\mu-n} \sum_{q=n-2\mu}^{n-\mu} b_{q-1}^{(n,n-\mu-1)}
 (A^{+} B^{+})^q (C^{+})^{2n-2\mu -2q+1} (Q^{+})^k \Psi_0 \\
{}= \alpha_0^{(k,n,2\mu+1} d_{k,n+1} (B^{+})^{2\mu-n} \sum_{q=n-2\mu}^{n-\mu} b_q^{(n+1,n-\mu)}
 (A^{+}B^{+})^q (C^{+})^{2n-2\mu-2q +1} (Q^{+})^k \Psi_0 \\
 \quad{} + \alpha_1^{(k,n,2\mu+1)} d_{k+1,n-1} (B^{+})^{2\mu-n} \Biggl[ 2 \sum_{q=n-2\mu+1}^{n-\mu}
 b_{q-1}^{(n-1,n-\mu-1)} (A^{+} B^{+})^q (C^{+})^{2n -2\mu-2q+1} \\
 \quad {} -\sum_{q=n-2\mu}^{n-\mu -1} b_q^{(n-1,n-\mu-1)} (A^{+}B^{+})^q (C^{+})^{2n-2\mu-2q +1}
 \Biggr] (Q^{+})^k \Psi_0,
\end{gather*}
thus leading to the following set of constraints by equating similar terms:
\begin{gather*}
 d_{k,n} b_{n-2\mu-1}^{(n,n-\mu-1)} =\alpha_0^{(k,n,2\mu+1)} d_{k,n+1} b_{n-2\mu}^{(n+1,n-\mu)} -
 \alpha_1^{(k,n,2\mu+1)} d_{k+1,n-1} b_{n-2\mu}^{(n-1,n-\mu-1)} \\
\hphantom{d_{k,n} b_{n-2\mu-1}^{(n,n-\mu-1)} =}{} \text{if $q=n-2\mu$}, \\
 d_{k,n} b_{q-1}^{(n,n-\mu-1)} =\alpha_0^{(k,n,2\mu+1)} d_{k,n+1} b_q^{(n+1,n-\mu)} \\
\hphantom{d_{k,n} b_{q-1}^{(n,n-\mu-1)} =}{}
 + \alpha_1^{(k,n,2\mu+1)}d_{k+1,n-1} \big(2 b_{q-1}^{(n-1,n-\mu-1)}-b_q^{(n-1,n-\mu-1)}\big)
\\
\hphantom{d_{k,n} b_{q-1}^{(n,n-\mu-1)} =}{}
\text{if $n-2\mu+1 \le q \le n-\mu-1$}, \\
 d_{k,n} b_{n-\mu-1}^{(n,n-\mu-1)} = \alpha_0^{(k,n,2\mu+1)} d_{k,n+1} b_{n-\mu}^{(n+1,n-\mu)} + 2
 \alpha_1^{(k,,n,2\mu+1)} d_{k+1,n-1} b_{n-\mu-1}^{(n-1,n-\mu-1)} \\
\hphantom{d_{k,n} b_{n-\mu-1}^{(n,n-\mu-1)} =}{} \text{if $q=n-\mu$}.
\end{gather*}
These can be shown to be satisfied by using the explicit formulas for $\alpha_0^{(k,n,m)}$, $\alpha_1^{(k,n,m)}$, and $b_q^{(n,n-\mu)}$. After considering the case where $n<2\mu+1$ in a similar way, the proof of the action of $A^+$ on the states $\Psi_{k,n,m}$ forming Jordan blocks is complete.

The action of $B^+$ and $C^+$ can be obtained in the same way by considering appropriate subcases for the index and deriving a set of constraints. We will only present the final results:
\begin{gather*}
 B^+\Psi_{k,n,m}=\beta_0^{(k,n,m)} \Psi_{k,n+1,m+2} +\beta_1^{(k,n,m)} \Psi_{k+1,n-1,m},
\end{gather*}
where
\begin{gather*}
 \beta_0^{(k,n,m)}=\frac{d_{k,n}}{d_{k,n+1}} \frac{(m+1)(m+2)}{2(n+1)(2n+1)}, \\
 \beta_1^{(k,nm)}=\frac{d_{k,n}}{d_{k+1,n-1}} \frac{2g^2n(2n-m-1)(2n-m)}{2n+1},
\end{gather*}
and
\begin{gather*}
 C^{+}\Psi_{k,n,m}=\gamma_0^{(k,n,m)} \Psi_{k,n+1,m+1} +\gamma_1^{(k,n,m)} \Psi_{k+1,n-1,m-1},
\end{gather*}
where
\begin{gather*}
 \gamma_0^{(k,n,m)}=-\frac{d_{k,n}}{d_{k,n+1}} \frac{m+1}{2g(n+1)(2n+1)}, \\
 \gamma_1^{(k,nm)}=\begin{cases}
\displaystyle \frac{d_{k,n}}{d_{k+1,n-1}} \frac{2gn(2n-m)}{2n+1} & \text{if $m \geq 1$}, \\
 0 & \text{if $m=0$}.
 \end{cases}
\end{gather*}

The action of the operators $A^-$, $B^-$, and $C^-$ is more complicated, because we have to use their commutation relations with $A^+$, $B^+$, $C^+$, and $Q^+$, given in Sections~\ref{section2} and~\ref{section3}, but it can be straightforwardly established. The results are given by
\begin{gather*}
 A^{-}\Psi_{k,n,m}=\bar{\alpha}_0^{(k,n,m)} \Psi_{k-1,n+1,m} +\bar{\alpha}_1^{(k,n,m)} \Psi_{k,n-1,m-2},
\end{gather*}
where
\begin{gather*}
 \bar{\alpha}_0^{(k,n,m)}=-\frac{d_{k,n}}{d_{k-1,n+1}} \frac{\lambda k}{g^2(n+1)(2n+1)}, \\
 \bar{\alpha}_1^{(k,n,m)}=\begin{cases}
 \displaystyle -\frac{d_{k,n}}{d_{k,n-1}} \frac{2\lambda n (2k +2n+1)}{2n+1} & \text{if $m \geq 2$}, \\
 0 & \text{if $m=0$ or 1},
 \end{cases}
\\
 B^{-}\Psi_{k,n,m}=\bar{\beta}_0^{(k,n,m)} \Psi_{k-1,n+1,m+1} +\bar{\beta}_1^{(k,n,m)}
 \Psi_{k-1,n+1,m+2} \nonumber \\
\hphantom{B^{-}\Psi_{k,n,m}=}{} +\bar{\beta}_2^{(k,n,m)} \Psi_{k,n-1,m-1} +\bar{\beta}_3^{(k,n,m)}
 \Psi_{k,n-1,m},
\end{gather*}
where
\begin{gather*}
 \bar{\beta}_0^{(k,n,m)}=\frac{d_{k,n}}{d_{k-1,n+1}} \frac{2k(m+1)}{(n+1)(2n+1)}, \\
 \bar{\beta}_1^{(k,n,m)}=-\frac{d_{k,n}}{d_{k-1,n+1}} \frac{2 \lambda k(m+1)(m+2)}{(n+1)(2n+1)}, \\
 \bar{\beta}_2^{(k,n,m)}=\begin{cases}
\displaystyle -\frac{d_{k,n}}{d_{k,n-1}} \frac{4 g^2 n (2n-m)(2k+2n+1)}{2n+1} & \text{if $m \geq 1$}, \\
 0 & \text{if $m=0$},
 \end{cases} \\
 \bar{\beta}_3^{(k,n,m)}=
 -\frac{d_{k,n}}{d_{k,n-1}} \frac{4\lambda g^2 n(2n-m)(2n-m-1)(2k+2n+1)}{2n+1},
\end{gather*}
and
\begin{gather*}
 C^{-}\Psi_{k,n,m}=\bar{\gamma}_0^{(k,n,m)} \Psi_{k-1,n+1,m} +\bar{\gamma}_1^{(k,n,m)}
 \Psi_{k-1,n+1,m+1} \nonumber \\
\hphantom{C^{-}\Psi_{k,n,m}=}{} +\bar{\gamma}_2^{(k,n,m)} \Psi_{k,n-1,m-2} +\bar{\gamma}_3^{(k,n,m)} \Psi_{k,n-1,m-1},
\end{gather*}
where
\begin{gather*}
 \bar{\gamma}_0^{(k,n,m)}=\frac{d_{k,n}}{d_{k-1,n+1}} \frac{k}{g(n+1)(2n+1)}, \\
 \bar{\gamma}_1^{(k,n,m)}=-\frac{d_{k,n}}{d_{k-1,n+1}} \frac{2 \lambda k(m+1)}{g(n+1)(2n+1)}, \\
 \bar{\gamma}_2^{(k,n,m)}=\begin{cases}
\displaystyle \frac{d_{k,n}}{d_{k,n-1}} \frac{2 g n (2k+2n+1)}{2n+1} & \text{if $m \geq 2$}, \\
 0 & \text{if $m=0$ or 1},
 \end{cases} \\
 \bar{\gamma}_3^{(k,n,m)} =\begin{cases}
\displaystyle \frac{d_{k,n}}{d_{k,n-1}} \frac{4\lambda g(2k+2n+1)n(2n-m)}{2n+1} & \text{if $m \ge 1$}, \\
 0 & \text{if $m=0$}.
 \end{cases}
\end{gather*}

Finally, let us consider the action of the ${\mathfrak{gl}}(3)$ linear Casimir operator, which, from (\ref{eq:casimir}), can be expressed in terms of the operators $H$, $R=A^+ A^-$, and $V=A^+C^-+C^+A^-$. From equation~(\ref{eq:associate}), it follows that
\begin{gather*}
 H \Psi_{k,n,m}= 2 \lambda (2k+n) \Psi_{k,n,m} + \Psi_{k,n,m-1},
\end{gather*}
while the results obtained above can be combined to yield
\begin{gather*}
 R \Psi_{k,n,m} =-\frac{d_{k,n}}{d_{k-1,n+2}} \frac{\lambda k}{4 g^4 ( n+1)(n+2)(2n+1)(2n+3)}
 \Psi_{k-1,n+2,m} \nonumber \\
\hphantom{R \Psi_{k,n,m} =}{} - \frac{\lambda (4k+2n +3)}{2g^2 (2n-1)(2n+3)} \Psi_{k,n,m-2} \nonumber \\
\hphantom{R \Psi_{k,n,m} =}{} -\frac{d_{k,n}}{d_{k+1,n-2}} \frac{2\lambda n (n-1)(2k+2n+1)}{(2n-1)(2n+1)} \Psi_{k+1,n-2,m-4}
\end{gather*}
and
\begin{gather*}
 V\Psi_{k,n,m} =\frac{d_{k,n}}{d_{k-1,n+2}} \frac{k}{4g^3 (n+1)(n+2)(2n+1)(2n+3)} \Psi_{k-1,n+2,m}
 \nonumber \\
\hphantom{V\Psi_{k,n,m} =}{} + \frac{4k+2n +3}{2g(2n-1)(2n+3)}\psi_{k,n,m-2}
 + \frac{\lambda}{g}\Psi_{k,n,m-1} \nonumber \\
\hphantom{V\Psi_{k,n,m} =}{} + \frac{d_{k,n}}{d_{k+1,n-2}} \frac{2g n (n-1)(2k+2n+1)}{(2n-1)(2n+1)} \Psi_{k+1,n-2,m-4}.
\end{gather*}
These equations lead to the following action for the linear Casimir operator
\begin{gather*}
 (E_{11}+E_{22}+E_{33})\Psi_{k,n,m}=\left(2k+n+\frac{3}{2}\right)\Psi_{k,n,m}.
\end{gather*}
We conclude that, in spite of the presence of Jordan blocks, the Casimir operator has the same action on their member states as on the eigenstates of the Hermitian harmonic oscillator.

\section{Construction of an extended biorthogonal basis}\label{section6}

The Hamiltonian being pseudo-Hermitian with $\eta=P_2$, one has to use a new scalar product \cite{bender05, bender07, mosta02a, mosta10}
\begin{gather*}
 \langle \Psi |\eta|\Phi \rangle=\langle\langle \Psi |\Phi\rangle\rangle= \int \Psi \Phi \,{\rm d}^3x.
\end{gather*}
For the ground state (\ref{eq:gs}), for instance, $\langle\langle \Psi_0|\Psi_0\rangle\rangle=\sqrt{\big(\frac{\pi}{\lambda}\big)^3}$. On the other hand, the nondiagonalizability of $H$ makes it necessary to add some associated functions to the wavefunctions in order to complete the Jordan blocks, thereby resulting in a set of functions $\Psi_{k,n,m}$, $k, n = 0, 1, \dots$, $m=0, 1, \dots, 2n$. With the corresponding functions $\tilde{\Psi}_{k,n,m}$ for $H^{\dagger}$, they make up an extended biorthogonal basis, whose scalar product is given by \cite{mosta02b, mosta02c}
\begin{gather*}
 \langle\langle \Psi_{k,n,m}|\Psi_{k',n',m'}\rangle\rangle = \langle \tilde{\Psi}_{k,n,m}|\Psi_{k',n',m'}\rangle =
 \int \Psi_{k,n,m} \Psi_{k',n',m'} \,{\rm d}^3x
 = \delta_{k,k'} \delta_{n,n'} \delta_{m,2n-m'}.
\end{gather*}

The purpose of this section is twofold: first, to determine the normalization coefficient $d_{k,n}$ of the functions $\Psi_{k,n,m}$ from the relation
\begin{gather}
 \langle\langle \Psi_{k,n,m} | \Psi_{k,n,2n-m} \rangle\rangle = 1 \label{eq:normalization}
\end{gather}
and second, to check the orthogonality of the functions $\Psi_{k,n,m}$ and $\Psi_{k,n,m'}$ with $m'\ne 2n-m$. In the calculations, we will rely on the commutation relations between $A^{\pm}$, $B^{\pm}$, $C^{\pm}$, $Q^{\pm}$ and use the following results:
\begin{gather*}
 \langle\langle {\cal O} \Psi_0 | {\cal O}' \Psi_0 \rangle\rangle= \langle \Psi_0 |{\cal O}^{\dagger} \eta
 {\cal O}' |\Psi_0 \rangle, \\
 (A^{+})^{\dagger} \eta =-\eta A^{-}, \qquad (B^{+})^{\dagger} \eta =-\eta B^{-}, \qquad
 (C^{+})^{\dagger} \eta =-\eta C^{-}, \qquad (Q^{+})^{\dagger} \eta =\eta Q^{-}, \\
 A^-\Psi_0 = B^-\Psi_0 = C^-\Psi_0 = Q^-\Psi_0 = 0.
\end{gather*}

\subsection{Calculation of the normalization coefficient}\label{section6.1}

We will first discuss the two special cases where $k=0$ or $n=0$, then consider the general case.

\subsubsection[Case k=0 and any n]{Case $\boldsymbol{k=0}$ and any $\boldsymbol{n}$}

The simplest calculation of $d_{0,n}$ corresponds to $m=0$ in (\ref{eq:normalization}). In such a case, we know from equations (\ref{eq:associate-1}) and (\ref{eq:c-d}) that
\begin{gather*}
 \Psi_{0,n,0} = d_{0,n} (2g)^{2n} n! (2n-1)!! (A^+)^n \Psi_0, \\
 \Psi_{0,n,2n} = d_{0,n} (B^+)^n \Psi_0.
\end{gather*}
On inserting these expressions in (\ref{eq:normalization}), it is easy to get the result
\begin{gather*}
 \langle\langle \Psi_{0,n,0} |\Psi_{0,n,2n} \rangle\rangle = d_{0,n}^2 \big(8g^2\lambda\big)^n (n!)^2 (2n-1)!!
 \langle\langle \Psi_0 | \Psi_0\rangle\rangle,
\end{gather*}
from which it follows that
\begin{gather}
 d_{0,n} = \left[2^{3n} (n!)^2 (2n-1)!! \lambda^n g^{2n} \left(\frac{\pi}{\lambda}\right)^{3/2}\right]^{-1/2}. \label{eq:d-0-n}
\end{gather}

As a check, we have proved that the conditions{\samepage
\begin{gather}
 \langle\langle \psi_{0,n,2\mu}|\psi_{0,n,2n-2\mu}\rangle\rangle = \langle\langle \psi_{0,n,2\mu+1}|
 \psi_{0,n,2n-2\mu-1} \rangle\rangle =1 \label{eq:normalization-bis}
\end{gather}
for any allowed $\mu$ value, lead to the same value for $d_{0,n}$ as given in~(\ref{eq:d-0-n}).}

Let us consider, for instance, $\langle\langle \psi_{0,n,2\mu}|\psi_{0,n,2n-2\mu}\rangle\rangle = 1$ for $n \geq 2\mu$ (or $n \leq 2n -2\mu$) and provide some key steps in the proof. From
\begin{gather*}
 \Psi_{0,n,2\mu}=d_{0,n} (B^{+})^{2\mu-n} \sum_{q=n-2\mu}^{n-\mu} a_q^{(n,n-\mu)} (A^{+} B^{+})^q
 (C^{+})^{2n-2\mu-2q} \Psi_0
\end{gather*}
and
\begin{gather*}
 \Psi_{0,n,2n-2\mu}=d_{0,n} (B^{+})^{n-2\mu} \sum_{r=0}^{\mu} a_r^{(n,\mu)} (A^{+}B^{+})^r
 (C^{+})^{2\mu-2r} \Psi_0,
\end{gather*}
we get
\begin{gather*}
 \langle\langle \Psi_{0,n,2\mu}|\Psi_{0,n,2n-2\mu}\rangle\rangle =d_{0,n}^2 (-1)^n
 \sum_{q=n-2\mu}^{n-\mu} \sum_{r=0}^{\mu} a_q^{(n,n-\mu)} a_r^{(n,\mu)} \nonumber \\
 \qquad{} \times \langle\Psi_0 | \eta (C^{-})^{2n-2\mu-2q} (B^{-})^{2\mu-n+q} (A^{-})^q (A^{+})^r
 (B^{+})^{n-2\mu+r} (C^{+})^{2\mu -2r}|\Psi_0 \rangle,
\end{gather*}
where
\begin{gather*}
\langle \Psi_0 | \eta (C^{-})^{2n-2\mu -2q} (B^{-})^{2\mu-n+q} (A^{+})^r (C^{+})^{2\mu-2r}
 \big[(A^{-})^{q},(B^{+})^{n-2\mu+r}\big]|\Psi_0\rangle \nonumber \\
\qquad =\begin{cases}
 0 & \text{if $q >n-2\mu +r$}, \\
\displaystyle (-2\lambda)^q \frac{(n-2\mu+r)!}{(n-2\mu+r-q)!} \langle \Psi_0|\eta (C^{-})^{2n-2\mu-2q}
 (B^{-})^{2\mu-n+q} & \\
 \quad \times (A^{+})^r (C^{+})^{2\mu-2r} (B^{+})^{n-2\mu+r-q} |\Psi_0\rangle & \text{if $q \leq
 n-2\mu+r$}.
 \end{cases}
\end{gather*}
This yields
\begin{gather*}
 \langle\langle \Psi_{0,n,2\mu}|\Psi_{0,n,2n-2\mu} \rangle\rangle = d_{0,n}^2 (-1)^n
 \sum_{q=n-2\mu}^{n-\mu} \sum_{r=q-n+2\mu}^{\mu} a_q^{(n,n-\mu)} a_r^{(n,\mu)} (-2\lambda)^q
 \frac{(n-2\mu+r)!}{(n-2\mu+r-q)!} \\
 \qquad {}\times\langle \Psi_0| \eta (C^{-})^{2n-2\mu-2q} (B^{+})^{n-2\mu +r -q} (B^{-})^{2\mu-n+q}
 (A^{+})^r (C^{+})^{2\mu -2r} |\Psi_0\rangle.
\end{gather*}
On using
\begin{gather*}
 (B^{-})^{2\mu -n +q} (A^{+})^r (C^{+})^{2\mu-2r} |\Psi_0 \rangle \nonumber \\
\qquad{} = \delta_{r,2\mu-n+q} \sum_{s=0}^{2\mu-n+q} 2^{2\mu -n +q} \begin{pmatrix} 2\mu - n + q \\ s
 \end{pmatrix} \frac{(2\mu -n +q)!}{s!} \frac{(2n-2\mu -2q)!}{(2n-2\mu -2q-s)!} \nonumber \\
\qquad \quad {}\times (-\lambda)^{2\mu -n +q-s} g^s (A^{+})^s (C^{+})^{2n-2\mu-2q-s} |\Psi_0\rangle
\end{gather*}
and
\begin{gather*}
 \langle\Psi_0|\eta (C^{-})^{2n-2\mu-2q} (A^{+})^s (C^{+})^{2n-2\mu-2q-s} |\Psi_0 \rangle \nonumber \\
 \qquad {}=\delta_{s,0} (-2\lambda)^{2n-2\mu-2q}(2n-2\mu-2q)! \langle\langle\Psi_0|\Psi_0\rangle\rangle,
\end{gather*}
we get
\begin{gather*}
 \langle\langle \Psi_{0,n,2\mu}|\Psi_{0,n,2n-2\mu}\rangle\rangle = d_{0,n}^2 \sum_{q=n-2\mu}^{n-\mu}
 a_q^{(n,n-\mu)} a_{2\mu-n+q}^{(n,\mu)} (2\lambda)^n \\
\hphantom{\langle\langle \Psi_{0,n,2\mu}|\Psi_{0,n,2n-2\mu}\rangle\rangle =}{}
 \times q! (2\mu -n +q)! (2n -2\mu - 2q)! \langle\langle \Psi_0 | \Psi_0 \rangle\rangle
\end{gather*}
and, with the explicit values of $a_q^{(n,n-\mu)}$ and $a^{(n,\mu)}_{2\mu-n+q}$,
\begin{gather*}
 \langle\langle \Psi_{0,n,2\mu}|\Psi_{0,n,2n-2\mu}\rangle\rangle = d_{0,n}^2 (2\lambda)^n (2g)^{2n} (n!)^2
 \mu! (n-\mu)! ( 2n-2\mu -1)!! (2\mu-1)!! \nonumber \\
 \qquad \times \sum_{q=n-2\mu}^{n-\mu}\frac{\Gamma(\frac{1}{2})}{q! (2\mu -n+q)! (n-\mu-q)!
 \Gamma(n-\mu-q+\frac{1}{2})}\langle\langle \Psi_0 |\Psi_0\rangle\rangle.
\end{gather*}
It only remains to use some standard binomial identity to transform the right-hand side of the latter equation into that of equation~(\ref{eq:normalization-bis}), which completes the proof for that case. The other cases can be dealt with in a similar way.

\subsubsection[Case n=0 and any k]{Case $\boldsymbol{n=0}$ and any $\boldsymbol{k}$}

In this case, equation (\ref{eq:normalization}) becomes $\langle\langle \Psi_{k,0,0}|\Psi_{k,0,0}\rangle\rangle = 1$, where
\begin{gather*}
 \langle\langle \Psi_{k,0,0}|\Psi_{k,0,0}\rangle\rangle =d_{k,0}^2 \langle \Psi_0|\eta (Q^{-})^k (Q^{+})^k |
 \Psi_0\rangle
 =d_{k,0}^2 \langle \Psi_0|\eta (Q^{-})^{k-1}\big[Q^{-},(Q^{+})^k\big] |\Psi_0 \rangle.
\end{gather*}
From the identity
\begin{gather}
 [Q^{-},(Q^{+})^k]=8k(Q^{+})^{k-1}\big[\lambda H-g V + (2k+1)\lambda^2\big]-16k(k-1)g^2 (Q^{+})^{k-2}
 (A^{+})^2, \label{eq:identity}
\end{gather}
which can be proved by induction over $k$, and the identities $H|\Psi_0\rangle =V|\Psi_0\rangle=0$, it follows that
\begin{gather*}
 \langle\langle \Psi_{k,0,0}|\Psi_{k,0,0}\rangle\rangle = d_{k,0}^2 \langle\Psi_0|\eta(Q^-)^{k-1}
 \big\{a^{(k)}_{1,0} (Q^+)^{k-1} + a^{(k)}_{1,1} (Q^+)^{k-2} (A^+)^2\big\}|\Psi_0\rangle,
\end{gather*}
with $a^{(k)}_{1,0} = 8k(2k+1)\lambda^2$ and $a^{(k)}_{1,1} = -16k(k-1)g^2$. Since $[Q^-, (A^+)^2]
|\Psi_0\rangle = -8\lambda A^+ A^- |\Psi_0\rangle = 0$, the latter equation can be rewritten as
\begin{gather*}
\langle\langle\Psi_{k,0,0}|\Psi_{k,0,0}\rangle\rangle = d_{k,0}^2 \langle \Psi_0|\eta (Q^-)^{k-2}
 \bigl\{
 a^{(k)}_{1,0} [Q^-, (Q^+)^{k-1}] + a^{(k)}_{1,1} [Q^-, (Q^+)^{k-2}] (A^+)^2\bigr\} |\Psi_0\rangle.
\end{gather*}
On using the identity (\ref{eq:identity}) again, we get
\begin{gather*}
\langle\langle\Psi_{k,0,0}|\Psi_{k,0,0}\rangle\rangle = d_{k,0}^2 \langle \Psi_0|\eta (Q^-)^{k-2}
 \nonumber \\
\qquad {}\times \bigl\{
 a^{(k)}_{2,0} (Q^+)^{k-2} + a^{(k)}_{2,1} (Q^+)^{k-3} (A^+)^2 + a^{(k)}_{2,2} (Q^+)^{k-4}
 (A^+)^4\bigr\} |\Psi_0\rangle
\end{gather*}
for some coefficients $a^{(k)}_{2,0}$, $a^{(k)}_{2,1}$, and $a^{(k)}_{2,2}$.

On pursuing in the same way, we arrive at the relation
\begin{gather*}
 \langle\langle \Psi_{k,0,0}|\Psi_{k,0,0}\rangle\rangle = d_{k,0}^2 \langle \Psi_0|\eta (Q^{-})^{k-l}
 \sum_{m=0}^{\min(l,k-l)} a_{l,m}^{(k)} (Q^{+})^{k-l-m} (A^{+})^{2m} |\Psi_0\rangle,
\end{gather*}
where $l=2$, 3, \dots, $k$ and $a_{l,m}^{(k)}$ are some coefficients. For $l=k$ we are left with
\begin{gather*}
 \langle\langle \Psi_{k,0,0}|\Psi_{k,0,0}\rangle\rangle = d_{k,0}^2 a_{k,0}^{(k)}
 \langle\langle \Psi_0 |\Psi_0\rangle\rangle,
\end{gather*}
where $a_{k,0}^{(k)}= a^{(k)}_{1,0} a^{(k-1)}_{1,0} \cdots a^{(1)}_{1,0} = 8^k k! (2k+1)!!\lambda^{2k}$. We have therefore proved that for any $k$
\begin{gather}
 d_{k,0}= \left[ 8^k k! (2k+1)!! \lambda^{2k} \left(\frac{\pi}{\lambda}\right)^{3/2}\right]^{-1/2}.
 \label{eq:d-k-0}
\end{gather}

\subsubsection[Case any k and any n]{Case any $\boldsymbol{k}$ and any $\boldsymbol{n}$}

Let us now turn ourselves to the general case and consider $m=0$ in equation (\ref{eq:normalization}). We plan to prove that provided the three following results
\begin{gather*}
 T_1 \equiv \langle \Psi_0|\eta (Q^{-})^{k-1} (A^{-})^n (A^{+})^2 (B^{+})^n (Q^{+})^{k-2} |\Psi_0\rangle =0, \\
 T_2 \equiv \langle \Psi_0|\eta (Q^{-})^{k-1} (A^{-})^n A^{+} (B^{+})^{n-1} (Q^{+})^{k-1} |\Psi_0\rangle =0, \\
 T_3 \equiv \langle \Psi_0|\eta (Q^{-})^{k-1} (A^{-})^n (B^{+})^{n-2} (Q^{+})^k |\Psi_0\rangle =0
\end{gather*}
are true, then
\begin{gather}
 \langle\langle \Psi_{k,n,0}|\Psi_{k,n,2n}\rangle\rangle = d_{k,n}^2 8^{k+n} 2^k (n!)^2 (2n-1)!! (1)_k
 \left(\frac{2n+3}{2}\right)_k g^{2n} \lambda^{2k+n} \langle\langle \Psi_0 |\Psi_0 \rangle\rangle,
 \label{eq:gen-norm}
\end{gather}
thus leading to the general expression
\begin{gather}
 d_{k,n} = \left[8^{k+n}k! (n!)^2 (2n+1)^{-1} (2n+2k+1)!! g^{2n} \lambda^{2k+n} \left(\frac{\pi}{\lambda}
 \right)^{3/2}\right]^{-1/2}. \label{eq:d-k-n}
\end{gather}
for the normalization coefficient.

We note that equation (\ref{eq:d-k-n}) agrees with equation (\ref{eq:d-0-n}) obtained for $k=0$ (as well as with equation (\ref{eq:d-k-0}) derived for $n=0$). Let us show that if equation (\ref{eq:d-k-n}) is valid for $k-1$ and any~$n$, and $T_1=T_2=T_3=0$, then it will be valid for $k$ and $n$.

On starting from
\begin{gather*}
 \Psi_{k,n,0}=d_{k,n} (2g)^{2n} n! (2n-1)!! (A^{+})^n (Q^{+})^k \Psi_0
\end{gather*}
and
\begin{gather*}
 \Psi_{k,n,2n}=d_{k,n} (B^{+})^n (Q^{+})^k \Psi_0,
\end{gather*}
we get
\begin{gather*}
 \langle\langle \Psi_{k,n,0}|\Psi_{k,n,2n}\rangle\rangle = d_{k,n}^2 (2g)^{2n} n! (2n-1)!! \langle \Psi_0 |
 \big[ (A^{+})^n (Q^{+})^{k-1}\big]^{\dagger} \eta Q^{-} (B^{+})^n (Q^{+})^k |\Psi_{0}\rangle.
\end{gather*}
The identity
\begin{gather*}
 [Q^{-},(B^{+})^n]=-4 n \lambda (B^{+})^{n-1} B^{-} -4 n g (B^{+})^{n-1} C^{-} -4 n (n-1) g^2 (B^{+})^{n-2},
\end{gather*}
which is easily proved by induction over $n$, leads to
\begin{gather*}
 Q^{-} (B^{+})^n (Q^{+})^k |\Psi_0\rangle =\big[ (B^{+})^n Q^{-} -4 n \lambda (B^{+})^{n-1} B^{-} -4 n g
 (B^{+})^{n-1} C^{-} \\
\qquad {}- 4n (n-1) g^2 (B^{+})^{n-2} \big](Q^{+})^k |\Psi_0\rangle
\end{gather*}
or
\begin{gather*}
 Q^{-} (B^{+})^n (Q^{+})^k |\Psi_0 \rangle \\
 \qquad{} =\big[ 8 \lambda^2 k (2n+2k+1) (B^{+})^n (Q^{+})^{k-1}
 -16 g^2 k (k-1) (A^{+})^2 (B^{+})^n (Q^{+})^{k-2} \nonumber \\
 \qquad\quad {}-16 g^2 k n A^{+} (B^{+})^{n-1} (Q^{+})^{k-1} - 4g^2 n (n-1) (B^{+})^{n-2} (Q^{+})^k \big]
 |\Psi_0\rangle,
\end{gather*}
where, in the last step, we used
\begin{gather*}
 Q^{-}(Q^{+})^k |\Psi_0\rangle = \big[ 8 \lambda^2 k(2k+1) (Q^{+})^{k-1} -16 g^2 k (k-1) (A^{+})^2
 (Q^{+})^{k-2}\big] |\Psi_0 \rangle, \\
 B^{-}(Q^{+})^k |\Psi_0\rangle = \big[{-}4 \lambda k B^{+} (Q^{+})^{k-1} -4 g k C^{+} (Q^{+})^{k-1}\big]|\Psi_0
 \rangle, \\
 C^{-}(Q^{+})^k |\Psi_0\rangle = \big[ 4 g k A^{+} (Q^{+})^{k-1} + 4 \lambda k C^{+} (Q^{+})^{k-1}\big]|\Psi_0
 \rangle,
\end{gather*}
also demonstrated by induction over $k$. This yields
\begin{gather*}
 \langle\langle\Psi_{k,n,0}|\Psi_{k,n,2n}\rangle\rangle \\
 \qquad{} =d_{k,n}^2 (2g)^{2n} n! (2n-1)!! \bigl\{8\lambda^2 k(2n+2k+1) \langle\Psi_0|\eta (Q^-)^{k-1}
 (A^-)^n (B^+)^n (Q^+)^{k-1}|\psi_0\rangle \\
 \qquad\quad{} -16g^2k(k-1)T_1 - 16g^2knT_2 - 4g^2n(n-1)T_3\bigr\} \\
 \qquad{} = d_{k,n}^2 (2g)^{2n} n! (2n-1)!! 8\lambda^2 k(2n+2k+1) \frac{\langle\langle\Psi_{k-1,n,0}|
 \Psi_{k-1,n,2n}\rangle\rangle}{d_{k-1,n}^2 (2g)^{2n} n! (2n-1)!!},
\end{gather*}
which amounts to equation (\ref{eq:gen-norm}) under the assumptions made.

\subsection{Orthogonality of the basis}\label{section6.2}

Let us start with the simplest case with $k=0$ and $n=1$. The corresponding Jordan block is spanned by the three states
\begin{gather*}
 \Psi_{0,1,0}=d_{0,1} 4 g^2 A^{+} \Psi_0,\qquad \Psi_{0,1,1}=d_{0,1} (-2g) C^{+} \Psi_0,\qquad \Psi_{0,1,2}
 =d_{0,1} B^{+} \Psi_0,
\end{gather*}
where use has been made of equations (\ref{eq:a}), (\ref{eq:b}), (\ref{eq:associate-2}), and
\begin{gather*}
 d_{0,1} = \left[8\lambda g^2\left(\frac{\pi}{\lambda}\right)^{3/2}\right]^{-1/2}
\end{gather*}
in accordance with equation (\ref{eq:d-0-n}). Direct calculations, as explained at the beginning of this section, lead to the following results
\begin{gather*}
 \langle\langle \Psi_{0,1,0}|\Psi_{0,1,0}\rangle\rangle = \langle\langle \Psi_{0,1,0}|\Psi_{0,1,1}\rangle\rangle
 = \langle\langle \Psi_{0,1,2}|\Psi_{0,1,2}\rangle\rangle =0,\\
 \langle\langle \Psi_{0,1,1}|\Psi_{0,1,2}\rangle\rangle=\frac{1}{2\lambda},
\end{gather*}
which show that the three states do not form an orthogonal basis, as it was expected. It is therefore necessary to orthogonalize it, which amounts to going from $\Psi_{0,1,m}$ to $\Phi_{0,1,m}$, defined by\looseness=-1
\begin{gather*}
 \Phi_{0,1,0}=\Psi_{0,1,0},\\
 \Phi_{0,1,1}=\Psi_{0,1,1}-\frac{1}{2\lambda} \Psi_{0,1,0},\\
 \Phi_{0,1,2}=\Psi_{0,1,2},
\end{gather*}
and such that $\langle\langle\Phi_{0,1,m}|\Phi_{0,1,m'}\rangle\rangle = \delta_{m,2-m'}$ for $m, m'= 0, 1, 2$.

We have studied in the same way some other Jordan blocks and constructed, from the $\Psi_{k,n,m}$ functions, new functions $\Phi_{k,n,m}$ satisfying the condition
\begin{gather*}
 \langle\langle\Phi_{k,n,m}|\Phi_{k',n',m'}\rangle\rangle = \delta_{k,k'} \delta_{n,n'} \delta_{m,2n-m'}.
\end{gather*}
For $k=0$ and $n=2$, for instance,
\begin{gather*}
 \Phi_{0,2,0} = \Psi_{0,2,0}, \\
 \Phi_{0,2,1} = \Psi_{0,2,1} - \frac{1}{2\lambda} \Psi_{0,2,0}, \\
 \Phi_{0,2,2} = \Psi_{0,2,2} + \frac{1}{6\lambda^2}\Psi_{0,2,0} - \frac{1}{2\lambda}\Psi_{0,2,1}, \\
 \Phi_{0,2,3} = \Psi_{0,2,3} + \frac{1}{48\lambda^3}\Psi_{0,2,0} - \frac{1}{24\lambda^2}\Psi_{0,2,1}, \\
 \Phi_{0,2,4}= \Psi_{0,2,4},
\end{gather*}
while, for $k=0$ and $n=3$,
\begin{gather*}
 \Phi_{0,3,0} = \Psi_{0,3,0}, \\
 \Phi_{0,3,1} = \Psi_{0,3,1} - \frac{1}{2\lambda} \Psi_{0,3,0}, \\
 \Phi_{0,3,2} = \Psi_{0,3,2} + \frac{3}{20\lambda^2} \Psi_{0,3,0} - \frac{1}{2\lambda} \Psi_{0,3,1}, \\
 \Phi_{0,3,3} = \Psi_{0,3,3} - \frac{1}{30\lambda^3} \Psi_{0,3,0} + \frac{3}{20\lambda^2} \Psi_{0,3,1}
 - \frac{1}{2\lambda} \Psi_{0,3,2}, \\
 \Phi_{0,3,4} = \Psi_{0,3,4} - \frac{1}{300\lambda^4} \Psi_{0,3,0} + \frac{1}{60\lambda^3} \Psi_{0,3,1}
 - \frac{1}{20\lambda^2} \Psi_{0,3,2}, \\
 \Phi_{0,3,5} = \Psi_{0,3,5}, \\
 \Phi_{0,3,6} = \Psi_{0,3,6}.
\end{gather*}

In this section, we have pointed out that the associated functions of the Jordan blocks for nondiagonalizable non-Hermitian Hamiltonians require much more involved calculations when going from two to three dimensions. The use of new ladder operators has, nevertheless, allowed us to conjecture the form of their general normalization coefficient and to enhance the need for an orthogonalization of the basis.

\section{Conclusion}\label{section7}

In the present paper, we have demonstrated that three sets of canonical ladder operators exist for the three-dimensional nonseparable and nondiagonalizable pseudo-Hermitian oscillator of \cite{barda}. They can be introduced from their action on the wavefunctions belonging to the lower lattice in Figure~\ref{fig1}.

These ladder operators have allowed us to show the existence of a nine-dimensional hidden symmetry algebra, which can be written in terms of ${\mathfrak{gl}}(3)$ generators. The latter can be expressed in terms of bosonic operators in a nonstandard realization, which serve to embed ${\mathfrak{gl}}(3)$ into an~${\mathfrak{sp}}(6)$ algebra, as well as into an ${\mathfrak{osp}}(1/6)$ superalgebra. Furthermore, we have connected the hidden symmetry algebra with the integrals responsible for the superintegrability of the model and established that the latter generate a cubic algebra.

The ladder operators have served to construct the associated functions completing the Jordan blocks, whose dimension has been established. We have also presented the action of these ladder operators on the associated functions, proved that the latter are eigenfunctions of the~${\mathfrak{gl}}(3)$ linear Casimir operator, and written a subset of associated functions as multivariate polynomials.

Finally, we have studied in detail the construction of an extended biorthogonal basis and shown that its structure is more complicated than that considered in~\cite{marquette20} for the corresponding two-dimensional model. Nevertheless, we have been able to conjecture the form of the associated function normalization coefficient and to establish the need for an orthogonalization of the basis.

The results obtained here point out some similarities between the present pseudo-Hermitian oscillator and the usual three-dimensional oscillator, but also indicate their very different nature. The method established in this paper may allow a broader understanding of these pseudo-Hermitian models. So far, only particular examples have been considered and no classification has been provided. Our understanding of the underlying hidden symmetry algebra may allow to provide other ideas to obtain and classify such models. Studies of supersymmetric quantum mechanics, superintegrability, separation of variables and hidden algebra have been restricted mostly to models whose states are eigenstates of the Hamiltonian. The present paper points out that models with Jordan blocks have features that make them interesting from a mathematical physics perspective. Among open problems, there is the application of those methods to pseudo-Hermitian anharmonic oscillator models \cite{cannata12}.

\appendix

\section[gl(3) generators in terms of ladder operators]{$\boldsymbol{{\mathfrak{gl}}(3)}$ generators in terms of ladder operators}\label{appendixA}

The purpose of this appendix is to list the expressions of the ${\mathfrak{gl}}(3)$ generators $E_{ij}$ in terms of the ladder operators $A^{\pm}$, $B^{\pm}$, and $C^{\pm}$:
\begin{gather*}
 E_{11}=-\frac{1}{2\lambda} C^{+}C^{-} + \frac{1}{2},\\
 E_{22}=\frac{\lambda}{2g^2} B^{+}B^{-}+\frac{1}{2\lambda}C^{+}C^{-}+\frac{1}{2g}(B^{+}C^{-}
 +C^{+}B^{-}) + \frac{1}{2}, \\
 E_{33}=-\frac{g^2}{2\lambda^3} A^{+}A^{-} -\frac{\lambda}{2g^2} B^{+}B^{-} -\frac{1}{2\lambda}C^{+}
 C^{-} -\frac{1}{2\lambda}(A^{+}B^{-}+B^{+}A^{-}) \nonumber \\
\hphantom{E_{33}=}{} -\frac{g}{2\lambda^2}(A^{+}C^{-}+C^{+}A^{-}) -\frac{1}{2g}(B^{+}C^{-}+C^{+}B^{-})
 + \frac{1}{2}, \\
 E_{12}={\rm i}\left( \frac{1}{2\lambda} C^{+}C^{-} + \frac{1}{2g} C^{+}B^{-}\right),\\
 E_{21}={\rm i} \left( \frac{1}{2\lambda}C^{+}C^{-}+ \frac{1}{2g} B^{+}C^{-}\right), \\
 E_{13}=-\frac{1}{2\lambda}C^{+}C^{-} -\frac{1}{2g} C^{+}B^{-} -\frac{g}{2\lambda^2}C^{+}A^{-}, \\
 E_{31}=-\frac{1}{2\lambda}C^{+}C^{-} -\frac{1}{2g} B^{+}C^{-} -\frac{g}{2\lambda^2}A^{+}C^{-}, \\
 E_{23}={\rm i}\biggl( \frac{\lambda}{2g^2}B^{+}B^{-}+\frac{1}{2\lambda}C^{+}C^{-} + \frac{1}{2\lambda}
 B^{+}A^{-} + \frac{1}{2g}B^{+}C^{-} + \frac{g}{2\lambda^2}C^{+}A^{-} + \frac{1}{2g} C^{+}B^{-}\biggr), \\
 E_{32}={\rm i}\biggl( \frac{\lambda}{2g^2}B^{+}B^{-}+\frac{1}{2\lambda}C^{+}C^{-} + \frac{1}{2\lambda}
 A^{+}B^{-} + \frac{1}{2g}C^{+}B^{-} + \frac{g}{2\lambda^2}A^{+}C^{-}
 + \frac{1}{2g} B^{+}C^{-}\biggr).
\end{gather*}

\section[Polynomials f\_q\textasciicircum{}\{n,p\}(u,w) for some low values of p and q]{Polynomials $\boldsymbol{f_q^{n,p}(u,w)}$ for some low values of $\boldsymbol{p}$ and $\boldsymbol{q}$}\label{appendixB}

The purpose of this appendix is to list some examples of polynomials $f_q^{n,p}(u,w)$, introduced in Section~\ref{section5.2}.

For $p=1$:
\begin{gather*}
\begin{split}
 q&=1, \qquad f_1^{(n,1)}=-4 g w, \\
 q&=2, \qquad f_2^{(n,1)}=-4 g^2.
\end{split}
\end{gather*}

For $p=2$:
\begin{gather*}
\begin{split}
 q&=1,\qquad f_{1}^{(n,2)}=-8 g^2 \lambda u, \\
 q&=2,\qquad f_{2}^{(n,2)}=16g^2 (w^2 -\lambda), \\
 q&=3,\qquad f_{3}^{(n,2)}=32 g^3 w, \\
 q&=4,\qquad f_{4}^{(n,2)}=16 g^4.
\end{split}
\end{gather*}

For $p=3$:
\begin{gather*}
 q=2,\qquad f_2^{(n,3)}=96 g^3 \lambda u w, \\
 q=3,\qquad f_3^{(n,3)}=-32g^3 \big({-}3 \lambda g u + 2 w^3 - 6 \lambda w \big), \\
 q=4,\qquad f_4^{(n,3)}=-192 g^4 \big(w^2 -\lambda \big), \\
 q=5,\qquad f_5^{(n,3)}=-192 g^5 w, \\
 q=6,\qquad f_6^{(n,3)}=-64 g^6.
\end{gather*}

\subsection*{Acknowledgments}

I.~Marquette was supported by Australian Research Council Future Fellowhip FT180100099. C.~Quesne was supported by the Fonds de la Recherche Scientifique - FNRS under Grant Number 4.45.10.08.

\pdfbookmark[1]{References}{ref}
\LastPageEnding


\begin{thebibliography}{99}
\footnotesize\itemsep=-0.6pt

\bibitem{bagchi}
Bagchi B.K., Supersymmetry in quantum and classical mechanics, \textit{Chapman
 \& Hall/CRC Monographs and Surveys in Pure and Applied Mathematics}, Vol.~116, Chapman \& Hall/CRC, Boca Raton, FL, 2001.

\bibitem{barda}
Bardavelidze M.S., Cannata F., Ioffe M.V., Nishnianidze D.N., Three-dimensional
 shape invariant non-separable model with equidistant spectrum,
 \href{https://doi.org/10.1063/1.4774292}{\textit{J.~Math. Phys.}} \textbf{54} (2013), 012107, 11~pages,
 \href{https://arxiv.org/abs/1212.4805}{arXiv:1212.4805}.

\bibitem{bender05}
Bender C.M., Introduction to ${\mathcal{PT}}$-symmetric quantum theory,
 \href{https://doi.org/10.1080/00107500072632}{\textit{Contemp. Phys.}} \textbf{46} (2005), 277--292,
 \href{https://arxiv.org/abs/quant-ph/0501052}{arXiv:quant-ph/0501052}.

\bibitem{bender07}
Bender C.M., Making sense of non-{H}ermitian {H}amiltonians, \href{https://doi.org/10.1088/0034-4885/70/6/R03}{\textit{Rep.
 Progr. Phys.}} \textbf{70} (2007), 947--1018, \href{https://arxiv.org/abs/hep-th/0703096}{arXiv:hep-th/0703096}.

\bibitem{bender98}
Bender C.M., Boettcher S., Real spectra in non-{H}ermitian {H}amiltonians
 having {${\mathcal{PT}}$} symmetry, \href{https://doi.org/10.1103/PhysRevLett.80.5243}{\textit{Phys. Rev. Lett.}} \textbf{80}
 (1998), 5243--5246, \href{https://arxiv.org/abs/physics/9712001}{arXiv:physics/9712001}.

\bibitem{bender01}
Bender C.M., Dunne G.V., Meisinger P.N., \d{S}im\d{s}ek M., Quantum complex
 {H}\'enon--{H}eiles potentials, \href{https://doi.org/10.1016/S0375-9601(01)00146-3}{\textit{Phys. Lett.~A}} \textbf{281} (2001),
 311--316, \href{https://arxiv.org/abs/quant-ph/0101095}{arXiv:quant-ph/0101095}.

\bibitem{cannata02}
Cannata F., Ioffe M.V., Nishnianidze D.N., New methods for the two-dimensional
 {S}chr\"odinger equation: {SUSY}-separation of variables and shape
 invariance, \href{https://doi.org/10.1088/0305-4470/35/6/305}{\textit{J.~Phys.~A: Math. Gen.}} \textbf{35} (2002), 1389--1404,
 \href{https://arxiv.org/abs/hep-th/0201080}{arXiv:hep-th/0201080}.

\bibitem{cannata10}
Cannata F., Ioffe M.V., Nishnianidze D.N., Exactly solvable nonseparable and
 nondiagonalizable two-dimensional model with quadratic complex interaction,
 \href{https://doi.org/10.1063/1.3298675}{\textit{J.~Math. Phys.}} \textbf{51} (2010), 022108, 14~pages,
 \href{https://arxiv.org/abs/0910.0590}{arXiv:0910.0590}.

\bibitem{cannata12}
Cannata F., Ioffe M.V., Nishnianidze D.N., Equidistance of the complex
 two-dimensional anharmonic oscillator spectrum: the exact solution,
 \href{https://doi.org/10.1088/1751-8113/45/29/295303}{\textit{J.~Phys.~A: Math. Theor.}} \textbf{45} (2012), 295303, 11~pages,
 \href{https://arxiv.org/abs/1206.4013}{arXiv:1206.4013}.

\bibitem{cooper}
Cooper F., Khare A., Sukhatme U., Supersymmetry and quantum mechanics,
 \href{https://doi.org/10.1016/0370-1573(94)00080-M}{\textit{Phys. Rep.}} \textbf{251} (1995), 267--385, \href{https://arxiv.org/abs/hep-th/9405029}{arXiv:hep-th/9405029}.

\bibitem{frappat}
Frappat L., Sciarrino A., Sorba P., Dictionary on {L}ie superalgebras,
 \href{https://arxiv.org/abs/hep-th/9607161}{arXiv:hep-th/9607161}.

\bibitem{genden}
Gendenshtein L.E., Derivation of exact spectra of the {S}chr\"odinger equation
 by means of supersymmetry, \textit{JETP Lett.} \textbf{38} (1983), 356--359.

\bibitem{ioffe}
Ioffe M.V., Cannata F., Nishnianidze D.N., Exactly solvable two-dimensional
 complex model with a real spectrum, \href{https://doi.org/10.1007/s11232-006-0092-7}{\textit{Theoret. and Math. Phys.}}
 \textbf{148} (2006), 960--967, \href{https://arxiv.org/abs/hep-th/0512110}{arXiv:hep-th/0512110}.

\bibitem{junker}
Junker G., Supersymmetric methods in quantum and statistical physics, \textit{Texts and
 Monographs in Physics}, \href{https://doi.org/10.1007/978-3-642-61194-0}{Springer-Verlag}, Berlin, 1996.

\bibitem{marquette20}
Marquette I., Quesne C., Ladder operators and hidden algebras for shape
 invariant nonseparable and nondiagonalizable models with quadratic complex
 interaction. {I}.~{T}wo-dimensional model, \href{https://doi.org/10.3842/SIGMA.2022.004}{\textit{SIGMA}} \textbf{18} (2022),
 004, 11~pages, \href{https://arxiv.org/abs/2010.15273}{arXiv:2010.15273}.

\bibitem{mosh}
Moshinsky M., Smirnov Yu.F., The harmonic oscillator in modern physics, Harwood,
 Amsterdam, 1996.

\bibitem{mosta02b}
Mostafazadeh A., Pseudo-{H}ermiticity for a class of nondiagonalizable
 {H}amiltonians, \href{https://doi.org/10.1063/1.1514834}{\textit{J.~Math. Phys.}} \textbf{43} (2002), 6343--6352,
 \href{https://arxiv.org/abs/math-ph/0207009}{arXiv:math-ph/0207009}.

\bibitem{mosta02a}
Mostafazadeh A., Pseudo-{H}ermiticity versus {$\mathcal{PT}$} symmetry: the
 necessary condition for the reality of the spectrum of a non-{H}ermitian
 {H}amiltonian, \href{https://doi.org/10.1063/1.1418246}{\textit{J.~Math. Phys.}} \textbf{43} (2002), 205--214,
 \href{https://arxiv.org/abs/math-ph/0107001}{arXiv:math-ph/0107001}.

\bibitem{mosta02c}
Mostafazadeh A., Erratum: {P}seudo-{H}ermiticity for a class of
 nondiagonalizable {H}amiltonians [\textit{J.~Math. Phys.} {\bf 43}, 6343 (2002)], \href{https://doi.org/10.1063/1.1540714}{\textit{J.~Math. Phys.}} \textbf{44} (2003), 943,
 \href{https://arxiv.org/abs/math-ph/0301030}{arXiv:math-ph/0301030}.

\bibitem{mosta10}
Mostafazadeh A., Pseudo-{H}ermitian representation of quantum mechanics,
 \href{https://doi.org/10.1142/S0219887810004816}{\textit{Int.~J. Geom. Methods Mod. Phys.}} \textbf{7} (2010), 1191--1306,
 \href{https://arxiv.org/abs/0810.5643}{arXiv:0810.5643}.

\bibitem{nana}
Nanayakkara A., Real eigenspectra in non-{H}ermitian multidimensional
 {H}amiltonians, \href{https://doi.org/10.1016/S0375-9601(02)01359-2}{\textit{Phys. Lett.~A}} \textbf{304} (2002), 67--72.

\bibitem{pauli43}
Pauli W., On {D}irac's new method of field quantization, \href{https://doi.org/10.1103/RevModPhys.15.175}{\textit{Rev. Modern
 Phys.}} \textbf{15} (1943), 175--207.

\bibitem{scholt92}
Scholtz F.G., Geyer H.B., Hahne F.J.W., Quasi-{H}ermitian operators in quantum
 mechanics and the variational principle, \href{https://doi.org/10.1016/0003-4916(92)90284-S}{\textit{Ann. Physics}} \textbf{213}
 (1992), 74--101.

\end{thebibliography}
\end{document}